\documentclass[11pt]{article}
\usepackage{amsmath,amssymb}
\usepackage[pdftex]{graphicx}
\usepackage{cite}
\begin{document}
\title{  Overlap integral of continuum stationary  states     }
\author{{Kenzo Ishikawa}$^{*(1,2)}$ and { Yuya Nishio}$^{*(1)}$} 
\maketitle
\begin{center}
(1)  Department of Physics, Faculty of Science, \\
Hokkaido University, Sapporo 060-0810, Japan \\
(2)Natural Science Center, \\
Keio University, Yokohama 223-8521, Japan\\
\end{center}


\begin{abstract}

The rigorous formula of overlap integrals of continuum stationary states  with  their asymptotic expressions 
 in potentials of finite widths are derived.   Those of   energies $E_1$ and $E_2$ consist  of diagonal terms that are proportional to $\delta(E_1-E_2)$ and nondiagonal terms.  Owing to the composition of nondiagonal terms,  superpositions of stationary states have  time-dependent norms and finite probability currents. These  do not represent isolate  states.    In various exceptional   potentials  and in free theory,   nondiagonal terms do not exist, and the superpositions  of   states with  different energies  represent isolate particles  that exactly describe scattering processes.

\end{abstract}
\newpage
\section{ Introduction }

Potential scattering is a typical phenomenon  of quantum mechanics and is  studied along with cross sections and other quantities that characterize a transition. 
The momentum dependence of a scattering rate  has been studied and represents  observable quantities  of short range origins\cite{goldberger,newton,taylor}. Recent experiments have demonstrated that the physical quantities of long range correlations  exist in quantum mechanics \cite{EPR,aspect}.  Determining transition probabilities  of isolate particles  including the long range correlations needs to be dealt with immediately.    The isolate states  are expressed by  superpositions of stationary states  satisfying 
orthogonality.

 Schr\"{o}dinger  equation determines the time-evolution of a wave function, and  
stationary solutions have a specific form in time $e^{\frac{E}{i\hbar }t}\psi_E(\vec x)$. Here $\psi_E(\vec x) $ is the eigenstate of the  Hamiltonian. These states supply physical quantities of the system. The  values obtained from exact solutions play a special role in general when these deviate from those  obtained from approximations.  
Stationary states of discrete energy levels are described by using $L^2$ functions, and Hilbert space methods are applied. Although the stationary states of continuous energy levels satisfy the same Schr\"{o}dinger  equation, they are not $L^2$ functions and Hilbert space methods are not applicable. Proofs that a set  of both functions form a complete set were given in \cite{newton,newton_p,coulomb_c}, nevertheless, the rigorous orthogonality of states with different energies remains uncertain.

Because stationary continuum   states  behave like free waves at the spatial infinity,  these are not normalized and those with different energies are approximately orthogonal \cite{Landau-Lifshits}. In the standard scattering method, the cross section is  computed from a probability current of stationary state. Mutual  non-orthogonality is irrelvant and  
  has  not been paid   serious attention. Violation of  orthonormalities  is in fact critical  for obtaining  rigorous transition probabilities of  isolate states. A probability current   flowing into the system  vanishes in the  isolate states. Conversely  a system of the finite current  is not isolated.   Now, a magnitude of the  net current  is proportional to a time-derivative of its norm from a conservation of a probability.
 The norm of the superpositions of  stationary states is constant if those with different energies are orthogonal, but is not so for non-orthogonal states.  
The orthogonality of stationary states is a key for physical applications, and is elucudated.   

   The  fact    that  the stationary continuum   states do not belong to a standard Hilbert space has inspired  various methods of creating   a self-adjoint extension of the  space, which are possible  by a quantization in a finite box,  a  regularized interaction $e^{-\epsilon |t|} H_{int}$, or   a  rigged Hilbert space. The first  two cases have been applied in many areas; \cite{Landau-Lifshits, 
Schiff,Iwanami-kouza,Lipkin}.   These methods represent idealistic physical systems  that are modified from the original ones. This point remains to be clarified if the scattering probability in experiments and  natural phenomena is  in complete agreement with these   idealistic  systems.   In the  method of a rigged Hilbert space,  an extension is not straightforward and practical calculations are not easy. As the outcome of the experiments lacks clarity, these points have not been studied in this paper.  
Existence of  scattering states and scattering operators under  the self-adjoint Hamiltonian has been studied  by   Kato, Simon, Kuroda, et al \cite{Kato,Simon,kuroda} by solving a time-dependent Schr\"{o}dinger equation perturbatively in the coordinate representation. A probability flux in  solutions represents an average probability current of stationary states, from which a cross section is defined.  The  cross section defined in this manner is an idealistic quantity for infinitely extended waves, which incompletely reflects  experimental situations of particles.

  In this paper, we provide a formula connecting the scalar product of continuum  states with asymptotic forms  of  wave functions, 
based on two physical requirements for scattering formalism:
$H_1$. Physical quantities are  scaling invariant in large system.
$H_2$. Isolate physical system is expressed by a state of  a constant norm.   
$H_1$ is applied to any physical systems but $H_2$ is applied to only isolate states.  For non-isolate states, $H_2$ is  unnecessary.   
We  clarify the issue including   the following concerns:   

$Q_1$.  Are   the Hamiltonian's eigenstates   orthogonal for  short-range potentials ?

$Q_2$. In what  potentials are the continuum states of different energies orthogonal ?

Energy eigenstates with continuous spectrum in specific potentials are known  to satisfy  orthogonality, whereas  in majority of short-range potentials the orthogonality   is not  satisfied. Despite of the proof of completeness, \cite{newton,newton_p,coulomb_c}, explicit expressions of scalar products were unknown.   
The orthogonality of states with different energies remains unclear despite of  fundamental   clue for the superpositions to provide  a correct scattering formalism shown in Appendix A-1.
Free waves and stationary states in exceptional potentials are orthogonal, and useful normalized states are constructed.   
These concerns  will prove beneficial when examining the  normalized states of many-body scatterings  also \cite{ishikawa-shimomura-PTEP, ishikawa-tobita-PTEP,ishikawa-tobita-ann, ishikawa-tajima-tobita-PTEP, maeda-PTEP,ishikawa-oda-PTEP,ishikawa-nishiwaki-oda-PTEP}.

The purpose of this   paper is to find overlap integrals  of stationary continuum  states with rigoruous solutions. We  clarify if their superposions are isolated using the overlap integrals.  The paper   is structured into the following different sections:  the first section provides an introduction and  the second section presents the fundamental properties  of stationary states. The third section provides formulae of  overlap integrals and  studies  
 orthogonality of the continuum  states in free theory  and in various specific potentials. These formulae  are  applied in systems of  general potentials. The fourth section provides proof of the nonorthogonality of exact solutions   for a square-well potential and other potentials. In the fifth  section, the  probability current of wave packets  are presented and  the sixth  section is devoted for a summary.
Appendix A provides a summary of earlier findings, while Appendix B is dedicated to regularization dependences.  
 
Henceforth, we will use a unit $\hbar =1$.

\section{Fundamental properties of stationary  states}
\subsection{Stationary states}

A  Schr\"{o}dinger time-dependent equation   
\begin{eqnarray}
& &i \frac{\partial}{\partial t} \psi(t,x) =H \psi(t,x) \label{Schoedinger}\\
& &H=H_0+V(x), \nonumber 
\end{eqnarray}
where $H_0=\frac{p^2}{2m}$, describes a particle of mass $m$ in potential $V(x)$.  In this system,  probability density function $\rho(\vec x,t)$ and the current $\vec j(\vec x,t)$ are defined as follows: 
\begin{eqnarray}
\rho(t,\vec x)&=&\psi^{*}(t,\vec x) \psi(t,\vec x), \label{density}\\
\vec j(t,\vec x)&=& \frac{1}{2m}( \psi^{*}(t,\vec x) ({-i \nabla} \psi(t,\vec x))-({-i \nabla}\psi^{*}(t, \vec x)) \psi(t,\vec x)). \label{current}
\end{eqnarray}
These satisfy   
\begin{eqnarray}
& &\frac{ \partial}{\partial t} \rho(t, \vec x)+ \nabla \cdot \vec j(t,\vec x)=0. 
\end{eqnarray}
Integrating over coordinate, we have the integral form 
\begin{eqnarray}
& &\frac{ d}{d t} N(t) + \int_{\partial V}  d \vec S \cdot \vec j(t,\vec x)=0, \label{conservation} \\
& & N(t)=\int_V  d \vec x \rho(t, \vec x),\label{conservation_i}
\end{eqnarray}
where $\partial V$ is a boundary and $N(t)$ is  norm.  

A special solution of the form   
\begin{eqnarray}
 \psi(t,x)=e^{\frac{Et}{i}} \phi(E,x),\label{separation}
\end{eqnarray}
represents a stationary state, which is unchanged with time.    
Substituting Eq.$(\ref{separation})$ with Eq.$(\ref{Schoedinger})$, we have 
\begin{eqnarray}
H\phi(E,x) =E \phi(E,x). 
\end{eqnarray} 
There are two kinds of states. One is bound state. Wave functions are normalized and  eigenvalues are discrete.  We write as  $E=E_l,l=0,1,\dots$. Scalar products of these functions satisfy
\begin{eqnarray}
\langle \phi(E_l,x)| \phi(E_{l'},x) \rangle =\int dx 
 \phi(E_l,x)^{*} \phi(E_{l'},x)=\delta_{l,l'}.\label{orthogonal_eigenstate_1}
\end{eqnarray}
Eq.$(\ref{orthogonal_eigenstate_1})$ is derived using the wave function to satisfy  the boundary condition $\phi(x)|_{x \rightarrow \infty}=0$. A general solution of Eq.$(\ref{Schoedinger})$ is expressed as follows:
 \begin{eqnarray}
& &\psi(t,x)= \sum_l  a(E_l) e^{\frac{E_lt}{i}} \phi(E_l,x). \label{superposition_b}
\end{eqnarray}
The norm of this  function
\begin{eqnarray}
& & \langle \psi(t,x)| \psi(t,x) \rangle =\sum_{l}  |a(E_l)|^2  
\end{eqnarray}
is constant with  time. 

The other is a continuum  state. Wave functions  behave as $e^{ikx}$ at $x= \pm \infty$, and the  state of the  wave number 
\begin{eqnarray}
\psi(t,x) =e^{\frac{E(k)t}{i}} \phi(E(k),x) \label{scattering_s}
\end{eqnarray}
has a time-independent norm. The first term of Eq.$( \ref{conservation})$ vanishes, and the cross section is computed from an expectation value of the probability current following an analogy with classical mechanics. The norm is given by 
\begin{eqnarray}
N=\int_{-\infty}^{\infty} dx |\phi(E(k),x)|^2
\end{eqnarray}
and  divergent. 

The proof that a  set of eigenfunctions $\phi(E_l,x)$ and $ \phi(E(k),x)$ form a complete set, \cite{newton,newton_p}, namely a function $h(x)$ of finite norm is expressed as  
\begin{eqnarray}
& &h(x)=\sum_l c_l \phi_l+ \int dk C(k) \phi(E(k),x) 
\end{eqnarray}
where
\begin{eqnarray}
& &c_l=\int_{-\infty}^{\infty} dx h(x)^{*} \phi_l(x),  \\
& &C(k)=\int_{-\infty}^{\infty}  dx h(x)     \phi^{*}(E(k),x), 
\end{eqnarray}
were  given long ago by Newton \cite{newton, newton_p} using perturbative expansions of Green's function, and  was extended to Coulomb potential later \cite{coulomb_c}. 
In these proofs, properties of Green's functions  under suitable boundary conditions were  used. These properties  are ensured with  Hermitian Hamiltonian, in which the orthogonality  of states with different energies are  satisfied. Nevertheless, the orthogonality of continuum states is unnecessary for these proofs, and is not satisfied in  rigorous solutions in short range potentials \cite{Landau-Lifshits}.  Orthogonality of states with different energies  remains  obscure.  

The orthogonality is crucial for the state to have a time independent norm.
Normalized  states consisting of  superpositions of waves  at $t=0$,
 \begin{eqnarray}
& &\psi(0,x)= h(x) \label{superposition_s}
\end{eqnarray}
   becomes at a later time $t$ 
 \begin{eqnarray}
& &\psi(t,x)= \sum_l c_l e^{\frac{E_l t}{i}} \phi_l+ \int dk C(k) e^{\frac{E(k)t}{i}} \phi(E(k),x).  \label{superposition_s2}
\end{eqnarray}
For  non-orthogonal states,
\begin{eqnarray}
\langle \phi(k,x)| \phi(k',x) \rangle \neq \delta(E(k)-E(k')) \label{orthogonal_eigenstate_s2},
\end{eqnarray}
 the norm of state  given by
\begin{eqnarray}
& &N(t)= \sum_l |c_l|^2 +\int dk dk'  a(E(k))^{*} a(E(k') ) e^{\frac{(E(k)-E(k'))t}{i}}   \langle \phi(k,x)| \phi(k',x) \rangle, \label{norm_scattering_s} \nonumber \\
& &
\end{eqnarray}
depends on time.
 $N(t)$ is a total probability  of the solution Eq.$(\ref{superposition_s2})$ and  varies with time. From Eq.$(\ref{conservation})$,  it is observed that this state  has a net  current flow of the probability.   Having a finite  current,  this state is connected with outside of the system, and   is  unsuitable for representing the isolate states.  
Studying scalar products of rigorous solutions carefully, we   are able to understand   these differences  and to clarify the issues for  physical  implications.      

Our object of this study is  to formulate a scattering amplitude and probability that is free from  this problem. 
\subsection{Scattering formalism and physical quantities}    
Cross sections  have been derived from  average  fluxes  of  stationary wave functions in the standard treatment based on time independent formalism, which correspond $(1A)$ in the table below, and based on time dependent formalism of $(2A)$ below.  These two methods provide the same results. Now, for the transition amplitude to include the long-range correlations, observables should be defined from the scalar product $\langle \psi_{out}|\psi_{in} \rangle $ of normalized initial and final states.  That is achieved in a time-dependent formalism $(2B)$. If the stationary states with different energies are orthogonal, the normalized waves can be  constructed and a corresponding formalism, which is denoted as $(1B)$, is possible in the time-independent formalism as well. The  orthogonality of the states is a key for this issue. 

Tables of scattering formalism of time-independent formalism:
\\
\begin{tabular}{|c|c|c|c|}
\hline
  formalism  & Wave functions & observable & boundary conditons  \\
 \hline
(1A) time-independent   & Non-normalized waves  & $\psi(x)^{*} \nabla \psi(x) $ &initial condition  \\
 \hline
  (1B) time-independent    & Normalized waves & $\langle \psi_{out}|\psi_{in} \rangle $  & initial and final conditions  \\
 \hline 
 \end{tabular}.
\\
\\
 
Tables of scattering formalism of time-dependent formalism:
\\
\begin{tabular}{|c|c|c|c|}
\hline
  formalism  & Wave functions & observable & boundary conditons  \\
 \hline
(2A) time-dependent   &Non-normalized waves  &$\psi(t,x)^{*} \nabla \psi(t,x) $&initial condition \\
 \hline
   (2B) time-dependent  &~ Normalized waves & $\langle \psi_{out}|\psi_{in} \rangle $ & initial and final conditions  \\
 \hline 
 \end{tabular}.
\\
\\
Observed physical quantities  are obtained from scalar product, $\langle \psi_{out} |\psi_{in} \rangle$,   or from probability current,  $\psi(x)^{*} \nabla \psi(x)$.

 \subsection{ Hermitian   Hamiltonian }

First, we studied the Hermitian property of Hamiltonian $H$ from a general viewpoint.

The integral of the potential term with the wave functions Eq.$(\ref{orthogonal_eigenstate_1})$ or Eq.$(\ref{scattering_s})$ satisfies 
\begin{eqnarray}
& &\int_{-\infty}^{\infty} dx \phi(E_2,x)^{*} V(x) \phi(E_1,x)= \int_{-\infty}^{\infty} dx ( V(x) \phi(E_2,x))^{*}  \phi(E_1,x)
\end{eqnarray}
for a short range potential. Equality  
\begin{eqnarray}
\int_{-\infty}^{\infty} dx  \phi(E_2,x)^{*} \frac{d^2}{dx^2}  \phi(E_1,x)= \int_{-\infty}^{\infty} dx  ( \frac{d^2}{dx^2}  \phi(E_2,x))^{*}  \phi(E_1,x)\label{hermitian_d}
\end{eqnarray}
 is satisfied for  bound state wave functions Eq.$(\ref{orthogonal_eigenstate_1})$ but it is not satisfied generally  for  scattering states Eq.$(\ref{scattering_s})$.


The behaviors of functions in the asymptotic regions influence the integrals, and the self-adjoint  property is sensitive to the values of the functions at $x=\pm \infty$ even in  
short-range potentials. 
 We will now study the orthogonality of stationary states from values of wave functions  at lower and upper limits of the integral. 
\subsubsection{Reduction of integrals for stationary continuum  states}
Eigenstate $\phi(k,x)  $ of an energy $E(k)=\frac{k^2}{2m}$ is complex and the complex conjugate   satisfies
\begin{eqnarray}
(H_0+V(x))  \phi(k,x)^{*}   = E(k)  \phi(k,x)^{*}. \label{stationary_2}
\end{eqnarray}
Multiplying wave functions with complex conjugates and subtracting them, we get  
\begin{eqnarray}
& &(H_0+V(x))  \phi(k_2,x)^{*} \phi(k_1,x)- \phi(k_2,x)^{*} (H_0+V(x))  \phi(k_1,x)   \nonumber \\
& &= (E(k_2)-E(k_1))  \phi(k_2,x)^{*} \phi(k_1,x) \label{stationary_3}.
\end{eqnarray}  
Using an identity   
 \begin{eqnarray}
& &  -\frac{1}{2m}( (\frac {d^2}{dx^2} \phi(k_2,x))^{*} \phi(k_1,x)-  \phi(k_2,x)^{*}\frac {d^2}{dx^2} \phi(k_1,x)) \nonumber \\
& &=-\frac{1}{2m}\frac{d}{dx}( (\frac {d}{dx} \phi(k_2,x))^{*} \phi(k_1,x)-  \phi(k_2,x)^{*}\frac {d }{dx} \phi(k_1,x)),  \label{stationary_4}
\end{eqnarray}  
we have an identity for integrals 
\begin{eqnarray}
& &-\frac{1}{2m}\int_{x_1}^{x_2} dx \frac{d}{dx}( (\frac {d}{dx} \phi(k_2,x))^{*} \phi(k_1,x)-  \phi(k_2,x)^{*}\frac {d }{dx} \phi(k_1,x)) \nonumber  \\
& &= (E(k_2)-E(k_1))  \int_{x_1}^{x_2} dx \phi(k_2,x)^{*} \phi(k_1,x),
\end{eqnarray}
for the scalar product of two continuum  states.
The scalar product is expressed by values at upper and lower limits  
\begin{eqnarray}
& &  \int_{x_1}^{x_2} dx \phi(k_2,x)^{*} \phi(k_1,x)= -\frac{1}{2m (E(k_2)-E(k_1)) }J(k_2,k_1)_{x_1}^{x_2},\label{comparison_integrals_1} \\
& &J(k_2,k_1)_{x_1}^{x_2}=( (\frac {d}{dx} \phi(k_2,x))^{*} \phi(k_1,x)-  \phi(k_2,x)^{*}\frac {d }{dx} \phi(k_1,x))|_{x_1}^{x_2}.  \label{comparison_integrals_2}
\end{eqnarray}
This identity is valid for arbitrary stationary states and lower and upper limits \cite{Messia}. Eqs.$(\ref{comparison_integrals_1})$ and   $( \ref{comparison_integrals_2})$ are useful identities. This relation would be  modified by a regularization.   
\subsubsection { Regularization at spatial infinity } 

For  regularized functions defined by an infinitesimal $\tilde \epsilon$,
\begin{eqnarray}
f_{reg}(x)=e^{-\tilde \epsilon |x|} f(x), g_{reg}(x)=e^{-\tilde \epsilon |x|} g(x), \label{regular}
\end{eqnarray}
where the limit $\tilde \epsilon \rightarrow 0$ is taken at the end. Asymptotic values vanish
\begin{eqnarray}
\lim_{x \rightarrow \pm \infty} f_{reg}(x)=e^{-\tilde \epsilon |x|} f(x)=0 ,\\
\lim_{x \rightarrow \pm \infty} g_{reg}(x)=e^{-\tilde \epsilon |x|} g(x)=0.
\end{eqnarray}
The regularized  functions vanish at the spatial infinity.

\section{Orthogonal  stationary  states}
The stationary states that have  different energies of continuous spectrum are orthogonal in free system, $\delta(x)$ potential, linear potential, and few  others. 
\subsection{ Plane waves }
A free wave  
\begin{eqnarray}
\phi(k,x)=e^{ikx}
\end{eqnarray}
of continious momentum has a scalar product given by  the left-hand side  of Eq.$(\ref{comparison_integrals_1})$ 
\begin{eqnarray}
& &  \int_{x_1}^{x_2} dx \phi(k_2,x)^{*} \phi(k_1,x)= \int_{x_1}^{x_2} dx e^{-i(k_2-k_1)x} = \frac{e^{-i(k_2-k_1)x_2} -e^{-i(k_2-k_1)x_1 }}{-i(k_2-k_1)}, \nonumber \\
& &
\end{eqnarray}
for  $k_1-k_2 \neq 0$ and the right-hand side  of Eq.$(\ref{comparison_integrals_1})$ is 
\begin{eqnarray}
 -\frac{1}{2 m(E(k_2)-E(k_1)) }(-i)(k_1+k_2)  ({e^{-i(k_2-k_1)x_2} -e^{-i(k_2-k_1)x_1 }}).
\end{eqnarray}
Hence, Eq.$(\ref{comparison_integrals_1})$ is satisfied. For $k_1-k_2=0$, a direct evaluation of  the left-hand side of Eq.$(\ref{comparison_integrals_1})$ agrees to the limit $k_1-k_2 \rightarrow 0$ of the right-hand side.    

For  $x_1=-\Lambda, x_2=\Lambda$, the scalar product approaches to delta function at $\Lambda \rightarrow \infty$,
\begin{eqnarray}
& &  \int_{-\Lambda}^{\Lambda} dx \phi(k_2,x)^{*} \phi(k_1,x)= \frac{e^{-i(k_2-k_1) {\Lambda} }-e^{i(k_2-k_1)\Lambda }}{-i(k_2-k_1)} \nonumber \\
& &=\frac{ -2i \sin  (k_2-k_1) {\Lambda}}{-i(k_2-k_1)} \rightarrow 2 \pi \delta(k_2-k_1). \label{dirac_delta}
\end{eqnarray}
{\bf Scaling hypothesis}

For a large $\Lambda$, physics of  scatterings  of $ x_1=-\Lambda/2, x_2=\Lambda/2$  is equivalent to that of $x_1=-\Lambda, x_2=\Lambda$. This scaling hypothesis is expressed by  
 \begin{eqnarray}
\frac{e^{-i(k_2-k_1) {\Lambda} }-e^{i  (k_2-k_1)\Lambda}}{-i(k_2-k_1)}=\frac{e^{-i(k_2-k_1) {\Lambda/2} }-e^{i  (k_2-k_1)\Lambda/2}}{-i(k_2-k_1)}. \label{scaling}
\end{eqnarray}
    Combining Eq.$(\ref{scaling})$ with an identity, 
\begin{eqnarray}
\frac{e^{-i(k_2-k_1) {\Lambda} }-e^{i  (k_2-k_1)\Lambda}}{-i(k_2-k_1)}=2\cos((k_2-k_1)\Lambda/2)\frac{e^{-i(k_2-k_1) {\Lambda/2} }-e^{i  (k_2-k_1)\Lambda/2}}{-i(k_2-k_1)},
\end{eqnarray}
we have   at $ \Lambda \rightarrow  \infty$ in   the scaling region,   
\begin{eqnarray}
2 \cos ((k_2-k_1)\Lambda/2) \rightarrow 1.
\end{eqnarray}
This  is solved  by 
\begin{eqnarray}
(k_2-k_1)\Lambda/2 \rightarrow \frac{\pi}{3} sign(k_1-k_2)+2\pi n
\end{eqnarray}
with an integer $n$. \footnote{Under this condition,  the large $\Lambda$ limit of the phase factor of the scalar product is not random and the scalar product becomes different from a result \cite{bar}.} At these momenta,   
\begin{eqnarray}
 \sin((k_2-k_1)\Lambda/2)\frac{e^{-i(k_2-k_1) {\Lambda/2} }-e^{i(k_2-k_1)\Lambda/2 }}{-i(k_2-k_1)} \nonumber \\
=  sign(k_2-k_1) \sin(\frac{\pi}{3})   \frac{e^{-i(k_2-k_1) {\Lambda/2} }-e^{i(k_2-k_1)\Lambda/2 }}{-i(k_2-k_1)}.
\end{eqnarray}
Values   at these momenta   depict   the  functional properties  in the scaling region Eq.$(\ref{scaling})$  despite $\delta(k_2-k_1)$ is not a normal function in Eq.$(\ref{dirac_delta})$.  Applying the same procedure, we have 
\begin{eqnarray}
& &\lim_{\Lambda \rightarrow \infty} \frac{e^{ - i(k_2-k_1) \Lambda}-1 }{ -i(k_2-k_1) } \rightarrow \lim_{\Lambda  \rightarrow \infty} e^{-i(k_2-k_1)(\Lambda/2)} \frac{e^{  -i(k_2-k_1) \Lambda/2}- e^{  i(k_2-k_1) \Lambda/2}}{- i(k_2-k_1)}  \nonumber \\
& & = \left( \cos (\pi/3)  -i \sin(\pi/3)  sign(k_2-k_1) \right)   2 \pi \delta(k_2-k_1). \label{dirac_delta_3}
\end{eqnarray}
Then we have,

{\bf Theorem 1}.
  
\begin{eqnarray}
& &\lim_{\Lambda \rightarrow \infty} \frac{e^{ - i(k_2-k_1) \Lambda}-1 }{ -i(k_2-k_1) } \rightarrow  \pi (1+i \sqrt 3 sign(k_1-k_2))\delta(k_2-k_1),  \label{delta_m} \nonumber \\
& &\lim_{\Lambda \rightarrow \infty }\frac{e^{i (k_2-k_1) \Lambda}-1 }{-i(k_2-k_1)} \rightarrow -\pi (1-i \sqrt 3 sign(k_1-k_2))\delta(k_2-k_1).
 \end{eqnarray}
The first  term in  the second line of above equation, Eq.$(\ref{dirac_delta_3})$,  is continuous but the second term is discontinuous in $k_2-k_1$ at $k_2-k_1=0$. An integral of its  product with continuous function over the momentum difference $k_2-k_1$ vanishes. Then the last term may be ignored in Eq.$(\ref{dirac_delta_3})$. The integrals in the scaling hypothesis  are used throughout this paper. See footnote  \footnote{ An expression of the left-hand side, $\frac{\cos (k_2-k_1) \Lambda -1}{-i(k_2-k_1)} +\frac{\sin(k_2-k_1) \Lambda}{k_2-k_1}$, is not proportional to $\delta(k_2-k_1)$ if $ \cos((k_2-k_1) \Lambda) \rightarrow 0 $ at $\Lambda \rightarrow \infty$ and is not appropriate.  }.
We should note that these momenta are different from those defined in a closed system that satisfy a periodic boundary condition.
Similarly for an integral  over a large interval $\Lambda$ and a lower limit $x_0$, a functional property  that depends on the interval is expressed as follows:  
\begin{eqnarray}
& &\lim_{\Lambda \rightarrow \infty} \frac{e^{ - i(k_2-k_1)  (\Lambda+x_0)}- e^{ - i(k_2-k_1) x_0}}{ -i(k_2-k_1)}=e^{-i(k_2-k_1)(x_0+\Lambda/2)} \nonumber \\
& &\times \frac{e^{  -i(k_2-k_1) \Lambda/2}- e^{  i(k_2-k_1) \Lambda/2}}{- i(k_2-k_1)} \rightarrow e^{-i(k_2-k_1)x_0}\pi \delta(k_2-k_1). \label{dirac_delta_2}
\end{eqnarray}
These  formulae ensure that  the delta function is defined by  the definite integral.  In the literature, other formula $\lim_{\Lambda \rightarrow \infty}  \cos \Lambda (k_2-k_1) =0$  is used sometimes \cite{sindelka}. However, at $k_2-k_1=0$, $  \cos \Lambda (k_2-k_1) =1$. 
Eqs.$(\ref{comparison_integrals_1})$,  $( \ref{comparison_integrals_2})$, $(\ref{dirac_delta})$,  $(\ref{delta_m})$,  and 
$(\ref{dirac_delta_2})$ are  key relations, and  used   throughout this paper.  

The integral of a square of the modulus over the momentum is given as
\begin{eqnarray}
\int d k_2 |\frac{ -2i \sin  (k_2-k_1) {\Lambda}}{-i(k_2-k_1)}  |^2= \Lambda \pi. 
\end{eqnarray}    
\subsubsection{ Superposition of  waves of opposite directions}
We confirm first Eqs.$(\ref{comparison_integrals_1})$ and $( \ref{comparison_integrals_2})$ for a wave
\begin{eqnarray}
\phi(k,x)=e^{ikx}+R(k) e^{-ikx}.
\end{eqnarray}
A scalar product is given by 
\begin{eqnarray}
& &\int_{x_1}^{x_2} dx \phi(k_2,x)^{*} \phi(k_1,x) \nonumber \\
& &=\frac{i}{k_2-k_1} (- (R(k_2)^{*} R(k_1) (e^{-i (k_1-k_2) x_2}- e^{-i (k_1-k_2) x_1})+e^{i(k_1-k_2)x_2} -e^{i(k_1-k_2)x_1}) \nonumber \\
& &+\frac{i}{k_1+k_2 } (R(k_1)( ( e^{-i(k_2+k_1)x_2}- e^{-i(k_2+k_1)x_1}) -R(k_2)^{*} ( e^{i(k_2+k_1)x_2} -  e^{i(k_2+k_1)x_1} )),\nonumber \\ 
& &\label{direct_2_1}
\end{eqnarray}and  the boundary value Eq.$(\ref{comparison_integrals_2})$ is given by  
\begin{eqnarray}
& &J(k_2,k_1)_{x_1}^{x_2}= \nonumber \\
& &-i(k_2+k_1)((e^{i(k_1-k_2)x_2} -e^{i(k_1-k_2)x_1})  -R(k_2)^{*}R(k_1)( e^{-i(k_1-k_2)x_2} - e^{-i(k_1-k_2)x_1} )) \nonumber \\
& &+i(k_1-k_2) (R(k_1)( e^{-i(k_1+k_2)x_2}-e^{-i(k_1+k_2)x_1} )  - R(k_2)^{*} ( e^{i(k_1+k_2)x_2} -e^{i(k_1+k_2)x_1}) ).  \nonumber \\
& & \label{intg_2} 
 \end{eqnarray}
We substitute this on the right-hand side of Eq.$( \ref{comparison_integrals_1} )$, and we have, 
for $x_1=-\Lambda,x_2=\Lambda$ and $\Lambda \rightarrow \infty$,using Eq.$(\ref{delta_m})$,
\begin{eqnarray}
 \int_{-\infty}^{\infty} dx \phi(k_2,x)^{*} \phi(k_1,x) &=& 2 \pi \delta(k_2-k_1) (R(k_2)^{*} R(k_1) +1 ) \nonumber \\
 &+& 2 \pi \delta(k_1+k_2) (R(k_1) +R(k_2)^{*} ).
\end{eqnarray}
Waves of $E_2 \neq E_1$ are orthogonal.
\subsection{ Short-range potential  } 
We next study continuum stationary states in a short-range potential $g \delta(x)$. Eigenfunctions for $m=1$ are
\begin{eqnarray}
\psi(k,x)&=&e^{ikx}+R(k) e^{-ikx} ;x<0 \label{delta_potential_1}\\
&=& T(k) e^{ikx} ;0<x,\nonumber 
\end{eqnarray}
where $R(k)$ and $T(k)$ are reflection and transmission  coefficients and 
\begin{eqnarray}
R(k)=\frac{-ig}{k+ig}, T(k)=\frac{k}{k+ig}\label{delta_potential_2}.
\end{eqnarray}
These satisfy
\begin{eqnarray}
|R(k)|^2+|T(k)|^2=1. \label{unit_flux}
\end{eqnarray}

\subsubsection{Reduced integrals}
Choose $x_1<0, x_2>0$  in Eqs.$(\ref{comparison_integrals_1})$ and $(\ref{comparison_integrals_2})$ for  general $R(k)$ and $T(k)$, and this  is expressed by  values at the upper and lower limits as follows:
\begin{eqnarray}
& &J(k_2,k_1)_{x_1}^{x_2} \nonumber\\
& &=-i(k_1+k_2)( (T(k_2)^{*} T(k_1) e^{i (k_1-k_2) x_2}- e^{i (k_1-k_2) x_1})+e^{-i(k_1-k_2)x_1} R(k_2)^{*}R(k_1)) \nonumber \\
& &-i(k_1-k_2) (R(k_1)e^{-i(k_1+k_2)x_1}- R(k_2)^{*} e^{i(k_1+k_2)x_1}).
\end{eqnarray}
We evaluate this at $x_2 \rightarrow \infty$ and $x_1 \rightarrow -\infty$ using a formula of a free wave Eq.$( \ref{delta_m} )$.
Substituting Eq.$(\ref{unit_flux})$, we have 

{\bf Theorem 2}.

The overlapp integral is decomposed to an energy conserving term and non-conserving term, and is expressed as,  

\begin{eqnarray}
& &-\frac{1}{k_2^2-k_1^2}J(k_2,k_1)_{x_1}^{x_2}\nonumber\\
&=&i( (T(k_2)^{*} T(k_1)e^{i (k_1-k_2) x_2}- e^{i (k_1-k_2) x_1}+e^{-i(k_1-k_2)x_1} R(k_2)^{*}R(k_1))\frac{-1}{k_1-k_2} \nonumber \\
& &-i{(R(k_1)e^{-i(k_1+k_2)x_1}- R(k_2)^{*} e^{i(k_1+k_2)x_1})}\frac{1}{k_1+k_2}\nonumber\\
&\rightarrow &  2  \pi \delta({k_1-k_2}) +{(R(k_1)+ R(k_2)^{*} )}  \pi \delta(k_1+k_2)-\Delta,\label{scalar_product}
\end{eqnarray}
where  
\begin{eqnarray}
& &\Delta=i( (T(k_2)^{*} T(k_1)-1) + R(k_2)^{*}R(k_1))\frac{1}{k_2-k_1} +i{(R(k_1)- R(k_2)^{*} )}\frac{1}{k_1+k_2}. \label{non-conserving} \nonumber \\
& &  
\end{eqnarray}
The first  and second terms in the right-hand side of Eq.$(\ref{scalar_product})$  vanish at $E_{k_1} \neq E_{k_2} $, whereas the third term does not vanish and represents energy non-conserving term.
From formulae  Eqs.$(\ref{scalar_product})$ and $(\ref{non-conserving})$,   the overlap integral is expressed by    the scattering coefficients, $R(k)$ and $T(k)$.  

$\Delta$ vanishes for $g\delta(x)$  from Eq.$(\ref{delta_potential_2})$. 
For a general short-range potential, $V(x)$, a wave function is expressed also as,
\begin{eqnarray}
\psi(k,x)&=&e^{ikx}+R(k) e^{-ikx} ;x<x_0 \label{delta_potential_3}\\
&=& T(k) e^{ikx} ;x_0'<x,\nonumber 
\end{eqnarray}
by different coefficients, $R(k)$ and $T(k)$. For $x_0 >x_1\rightarrow -\infty$ and $x_0' < x_2  \rightarrow \infty$, Eq.$(\ref{scalar_product})$ is valid.  $\Delta$ is expressed by expression, 
Eq.$(\ref{non-conserving} )$, and
  does not  vanish at $E(k_1) \neq E(k_2)$. 
\subsubsection{Regularized integrals}
We study a case of regularized integrals.  By applying Eq.$(\ref{regular})$, we get  the integral  
\begin{eqnarray}
& &\langle \phi(k_2,x)| \phi(k_1,x) \rangle \nonumber  \\
& &=\int_{-\infty}^{0} dx e^{(i(-k_2+{k_1})+\epsilon )x} +\int_{0}^{\infty} dx e^{(i(-k_2+{k_1})-\epsilon) x}(T(k_2)^{*}T(k_1)+R(k_2)^{*}R(k_1)) \nonumber \\
& &+\int_{-\infty}^{0} dx R(k_1) e^{(-i(k_1+{k_2})+\epsilon)x} + \int_{-\infty}^{0} dx R(k_2)^{*}  e^{(i(k_1+{k_2})+\epsilon)x} \nonumber\\
& &=\pi \delta(k_2-k_1)( 1+T(k_2)^*T(k_1)+R(k_2)^*R(k_1))+\pi\delta(k_1+k_2))( R(k_1)+R(k_2)^*) \nonumber\\
& & +i P\frac{1}{k_2-k_1} (1- T(k_2)^{*}T(k_1)-R(k_2)^{*}R(k_1)))+i P\frac{1}{k_1+k_2} ( R(k_1)-R(k_2)^* ). \nonumber \\
& &
\end{eqnarray}
Substituting Eq.$(\ref{delta_potential_2})$, we have the second term on the right-hand side
\begin{eqnarray}
g\pi\delta(k_1+k_2)( \frac{ik_1-g-ik_2-g }{(k_1+ig)(k_2-ig)}),
\end{eqnarray}
and the  sum of third and fourth terms 
\begin{eqnarray}
& & +i(-i) P\frac{1}{k_2-k_1}  \frac{k_2 -k_1 }{(k_2-ig)(k_1+ig)}) +i^2 P\frac{1}{k_1+k_2} ( \frac{k_1+k_2}{(k_1+ig)(k_2-ig)} ), \label{cancel} \nonumber \\
& &
\end{eqnarray}
which   can be  expressed as follows: 
\begin{eqnarray}
& & +i(-i) ( P\frac{1}{k_2-k_1} (k_2-k_1)  - P\frac{1}{k_1+k_2}(k_1+k_2))  \frac{1 }{(k_2-ig)(k_1+ig)}   =0. \nonumber \\ \label{cancel_1}
\end{eqnarray}
This result  is consistent with value Eq.$(\ref{non-conserving})$.  We should note that another expression of Eq.$(\ref{cancel})$  using infinitesimal $\epsilon$ is the following:
\begin{eqnarray}
& & ( \frac{(k_2-k_1)^2}{(k_2-k_1)^2+\epsilon^2} - \frac{(k_1+k_2)^2}{(k_1+k_2)^2+\epsilon^2}  )\frac{1 }{(k_2-ig)(k_1+ig)},  \nonumber \\
\end{eqnarray}
which vanishes in the region $\epsilon \ll |k_1 \pm k_2|$ but does not vanish in other regions.  

Finally we have,
\begin{eqnarray}
& &\langle \phi(k_2,x)| \phi(k_1,x) \rangle =g(2\pi) \delta(k_2-k_1) +\pi\delta(k_1+k_2)( \frac{ik_1-g-ik_2-g }{(k_1+ig)(k_2-ig )}). \nonumber \\
& &
\end{eqnarray}
$\Delta$ vanishes. Stationary states are orthogonal for the energies $E(k_1) \neq E(k_2)$ in the present regularizations. 


 The stationary scattering states of different energies are orthogonal in the following potentials.

\subsection{Uniform  force }
Stationary states in a linear potential are extended of continuous spectrum. Such case are a massive  particle in a uniform gravity potential or a charged particle in a uniform electric field.   Here, a massive particle of mass $m$ in a uniform gravity is studied. A Hamiltonian  is
\begin{eqnarray}
H=\frac{ p_z^2}{2m}+mg z,   
\end{eqnarray}
and an eigenvalue equation  
\begin{eqnarray}
 (\frac{ p_z^2}{2m}+mg z) \phi_E=E\phi_E
\end{eqnarray}
is solved with     
\begin{eqnarray}
& &\phi_E(z)=A_E (\xi-\xi_1), \label{airy_solution}\\
& &c \xi=z, c\xi_1=  \frac{E}{mg}, c=(\frac{1}{2m^2 g})^{1/3}.\nonumber 
\end{eqnarray}
$ A_E (\xi-\xi_1)$ is    Airy function defined by  
\begin{eqnarray}
A_E (\xi)= \frac{1}{\pi}\int_0^{\infty} du \cos (\frac{u^3}{3} +\xi u), 
\end{eqnarray}
where $\xi_1$ shows a position of a turning point, and it depends on $E$. \cite{Landau-Lifshits}.

Scalar products of Airy functions are as follows: 
\begin{eqnarray}
& &\int_{-\infty}^{\infty} dtA_i(t+x) A_i(t+y) \nonumber\\
& &=\int_{-\infty}^{\infty} dt \frac{1}{\pi} \int_0^{\infty} du\cos( \frac{u^3}{3}+(t+x)u) \frac{1}{\pi} \int_0^{\infty} dv \cos (\frac{v^3}{3}+(t+y)v) \nonumber \\
& &=\int_{-\infty}^{\infty} dt \frac{1}{\pi^2} \int_0^{\infty} \frac{du dv}{2} [ \cos (\frac{u^3+v^3}{3} +(t+x)u+(t+y)v) \nonumber \\
& &~~~+ \cos (\frac{u^3-v^3}{3} +(t+x)u-(t+y)v) ].
\end{eqnarray}
Interchanging the order of integrations,   we have  
\begin{eqnarray}
& &\int_{-\infty}^{\infty} dtA_i(t+x) A_i(t+y)= \frac{1}{\pi^2} \int_0^{\infty} \frac{du dv}{2} [ \cos (\frac{u^3+v^3}{3} +xu+yv)2 \pi \delta( u+v) \nonumber \\
& &~~~+ \cos (\frac{u^3-v^3}{3} +xu-yv) (2 \pi) \delta(u-v)] \nonumber\\
& &= \frac{1}{\pi^2} \int_0^{\infty} \frac{du }{2} [ \cos  (x-y)u  (2 \pi) ] \nonumber\\
& &=\delta(x-y),
\end{eqnarray}
and 
\begin{eqnarray}
& &\int_{-\infty}^{\infty} dtA_i(t+x) \frac{\partial}{\partial t} A_i(t+y) \nonumber\\
& &=\frac{\partial}{\partial y} \int_{-\infty}^{\infty} dtA_i(t+x) A_i(t+y) \nonumber\\
& &=\frac{\partial}{\partial y}\delta(x-y).
\end{eqnarray}
Although  wave functions $(\ref{airy_solution})$ are totally different from plane  waves, their scalar products are proportional to delta functions. 

\subsection{Confining potential in three dimensions}
A potential between quarks in mesons has a form
\begin{eqnarray}
V(r)=kr +V_s(r), V_s(r)|_{r =\infty} \rightarrow 0.
\end{eqnarray} 
This potential satisfies 
\begin{eqnarray}
\lim_{r \rightarrow \infty} V(r)= \infty,
\end{eqnarray}
 and all  energy eigenstates are bound states described by $L^2$
functions. Accordingly stationary states 
are bound states and  satisfy  
\begin{eqnarray}
& & \lim_{r \rightarrow \infty} \psi(\vec  x)=0 ,\nonumber \\
& &\int d{\vec x} \psi_{E_i}( \vec x)^{*} \psi_{E_j} ( \vec x)=\delta_{i,j}, \text{for} ~E_i \neq E_j.
\end{eqnarray}
\subsection{Harmonic oscillator}
For a harmonic oscillator, the potential is
\begin{eqnarray}
V(x)=\frac{k}{2} x^2,
\end{eqnarray}
which becomes $\infty$ at $x \rightarrow \pm \infty$.  All  energy eigenstates are bound states described by $L^2$
functions. 
\subsection{ Landau levels}
In the following gauge choice, two-dimensional electrons in the perpendicular magnetic field are  equivalent to a plane wave in one direction and a harmonic oscillator in perpendicular direction in the following gauge choice,
\begin{eqnarray}
(A_x,A_y)  =(0,x) B.
\end{eqnarray}
The Hamiltonian is given as
\begin{eqnarray}
H=\frac{p_x^2+(p_y+eBx)^2}{2m},
\end{eqnarray}
which is the sum of the harmonic oscillator in the $x$ direction and a free particle  in the $y$ direction. Eigenstates are products of the plane wave and harmonic oscillator eigenstates $ e^{ik_y y} H_l(x-x_o(k_y))$. These satisfy  orthogonality. In different gauges, wave functions become different of satisfying  orthogonality. \cite{Landau-level}

\subsection{Periodic  potentials}
Eigenstates in a periodic potential are expressed by plane waves, and the two states with  different energies are orthogonal. Energy eigenvalues are different from those of the free system, but  the orthogonality of stationary states is  almost equivalent to that of  free waves. These   are expressed by using  Bloch's theorem. \cite{Bloch}
\subsection{Coulomb  potential in three dimensions}
\subsubsection{ Bound states}
A potential between electric charges has a form
\begin{eqnarray}
V(r)=k\frac{1}{r} , 
\end{eqnarray} 
and satisfies 
\begin{eqnarray}
\lim_{r \rightarrow \infty} V(r)= 0.
\end{eqnarray}
Bound  states exist for $k<0$ and satisfy  orthonormality 
\begin{eqnarray}
\int d{\vec x} \psi_{E_i}( \vec x)^{*} \psi_{E_j} ( \vec x)=\delta(E_i-E_j).
\end{eqnarray}

\subsubsection{Continuum  states}
Scattering states for $k<0$ and ststes for $k>0$ are expressed  with hyperbolic coordinates $(\xi,\eta,\phi)$ as
\begin{eqnarray}
\left( \frac{4}{\xi+\eta} [\frac{\partial }{\partial \xi}(\xi \frac{\partial}{\partial \xi} + \frac{\partial }{\partial \eta }(\eta  \frac{\partial}{\partial \eta }  ] +\frac{1}{\xi \eta} \frac{\partial^2}{{\partial \phi}^2}+\epsilon+\frac{4 m \alpha}{\hbar^2 (\xi+\eta)} \right) \psi=0,
\end{eqnarray} 
where 
\begin{eqnarray}
\epsilon=\frac{p_{\infty}^2}{2m}.
\end{eqnarray}
Axial symmetric solution
\begin{eqnarray}
\psi=C_x e^{i p_{\infty}\frac{\xi-\eta}{2 \hbar}} \tilde Y(\eta)
\end{eqnarray}
where $\tilde Y(\eta)$ is confluent hypergeometric function  
\begin{eqnarray}
& &\eta \frac{d^2 \tilde Y}{ d \eta^2}+(1-\frac{i p_{\infty}}{\hbar} \eta) \frac{d \tilde Y}{d \eta} +\frac{ \gamma p_{\infty}}{\hbar} \tilde Y=0 \\
& & \gamma=\frac{ m \alpha}{\hbar p_{\infty}}.
\end{eqnarray} 
Volume element is 
\begin{eqnarray}
dx dy dz= (\frac{1}{2})^2 (\xi+\eta) d \xi d \eta  d \phi.
\end{eqnarray}
Scalar product
\begin{eqnarray}
& &(\psi_{\epsilon}, \psi_{\epsilon'})=\int (\frac{1}{2})^2 (\xi+\eta) d \xi d \eta  d \phi { \psi_{\epsilon}(\xi,\eta,\phi)}^{*} \psi_{\epsilon'}(\xi,\eta,\phi) 
\end{eqnarray}
for $\phi$ independent functions is defined with variables $\eta$ and $\xi-\eta$ as 
\begin{eqnarray}
& &\int (\frac{1}{2})^3 (\xi-\eta+2 \eta) d {\eta } \left( \int d ({\xi-\eta} ) d \phi { \psi_{\epsilon}(\xi,\eta,\phi)}^{*} \psi_{\epsilon'}(\xi,\eta,\phi)|_{\eta}\right) \nonumber \\
& &=\int (\frac{1}{2})^3  d {\eta } \left( \int  (\xi-\eta)d ({\xi-\eta} ) e^{-i( p_{\infty}-p_{\infty}')\frac{\xi-\eta}{2 \hbar}}d \phi {\tilde Y_{p_{\infty}}(\eta)}^{*} \tilde Y_{p_{\infty}'}(\eta)|_{\eta}\right) \nonumber \\
& &+\int (\frac{1}{2})^3 (2 \eta) d {\eta } \left( \int d ({\xi-\eta} )e^{-i( p_{\infty}-p_{\infty}')\frac{\xi-\eta}{2 \hbar}} d \phi  {\tilde Y_{p_{\infty}}(\eta)}^{*} \tilde Y_{p_{\infty}'}(\eta)|_{\eta} \right) \nonumber \\
& &=\int (\frac{1}{2})^3  d {\eta } \left( 4 \pi \hbar \delta'(p_{\infty}-p_{\infty}')d \phi {\tilde Y_{p_{\infty}}(\eta)}^{*} \tilde Y_{p_{\infty}'}(\eta)|_{\eta}\right)\nonumber \\
& &+\int (\frac{1}{2})^3 (2 \eta)  \left( 4 \pi \hbar \delta(p_{\infty}-p_{\infty}') 
   \int d \eta d \phi  {\tilde Y_{p_{\infty}}(\eta)}^{*} \tilde Y_{p_{\infty}'}(\eta)|_{\eta} \right). \nonumber \\
& &= (\frac{1}{2})^3    4 \pi \hbar \delta'(p_{\infty}-p_{\infty}')(2 \pi) \int d {\eta }
{\tilde Y_{p_{\infty}}(\eta)}^{*} \tilde Y_{p_{\infty}'}(\eta)\nonumber \\
& &+ (\frac{1}{2})^3    4 \pi \hbar \delta(p_{\infty}-p_{\infty}') 
  (2 \pi)  \int d {\eta }  (2 \eta){\tilde Y_{p_{\infty}}(\eta)}^{*} \tilde Y_{p_{\infty}'}(\eta). 
\end{eqnarray}
Then the scalar product of scattering states with energies $E_1$ and $E_2$ is composed of $\delta(E_1-E_2)$ and $\delta'(E_1-E_2)$ and vanish for $E_1 \neq E_2$.   

\subsection{Summary of  orthogonal stationary states}
The stationary states with energies $E_1$ and $E_2$  of this section except Coulomb potential  are  $\delta(E_1-E_2)$.  The superpositions of states with  different energies  have constant norms. For  Coulomb potential, the scalar product  has a term proportional to $\delta'(E_1-E_2)$ in addition to $\delta(E_1-E_2)$.        
The overlap integrals  of plane waves satisfies $\bf Theorem ~1$, which  is a basic formula of the Fourier transformation. This 
 is  used as a standard formula and $ \bf Theorem ~2$ holds for other cases. 
For long-range  potentials, as described in  this section, wave functions show the orthogonality.   Potential  $\delta(x)$ is  exceptional among short-range potentials, and  stationary scattering states with  different energies are orthogonal.   
 
\section{  Overlap integrals of non-orthogonal stationary states   }
Stationary continuum  states in   square-well potential and other short-range potentials  are studied. Wave functions are plane waves at spatial infinity, and modified  at potential regions. Consequently the scalar products of states with  different energies are given  by the  sum of  Dirac delta functions and non-orthogonal terms. For the square-well potential  solutions are known, but the scalar products are not found in the literture.  Calculations are presented.

\subsection{ Square-well potential : energy eigenstates }
 
We study a particle of mass $m=1$ in a parity-symmetric potential: 
\begin{eqnarray}
V(x)&=&0, ~~x \leq -a/2\nonumber \\ 
&=&V_0, ~~-a/2<  x \leq a/2\nonumber \\
&=&0  ~~a/2 <x. \label{square_well}
\end{eqnarray}
 Stationary scattering states and their first derivatives are continuous at boundaries $x=\pm a/2$.    

For  bound states, 
\begin{eqnarray}
\phi_m^{*}= \phi_m
\end{eqnarray}
and 
\begin{eqnarray}
\int dx \phi_m^{*} \phi_n =\delta_{m,n}.
\end{eqnarray}
Eigenstates of continuum spectra are expressed as
\begin{eqnarray}
\phi(k,x)& =&e^{ikx}+R(k) e^{-ikx}; x \leq - a/2 \nonumber \\
&=&  A_{+} (k)e^{i{\hat k}x}+A_{-}(k) e^{-i{\hat k}x}; -a/2<x \leq  a/2 \nonumber \\
&=& T(k) e^{ikx}; a/2<x, \label{eigenstates_sp}
\end{eqnarray}
where 
\begin{eqnarray}
E=\frac{k^2}{2}=\frac{{\hat k}^2}{2}+V_0.
\end{eqnarray}
We study a case of real $k$ and $\hat k$ that is as follows. Other cases were  also studied  similarly.  
The boundary conditions are the continuity of the function and the first derivative at $x= - a/2$
\begin{eqnarray}
& &e^{-ika/2}+R(k) e^{ika/2}=A_{+}(k) e^{-i{\hat k}a/2}+A_{-}(k) e^{i{\hat k}a/2}, \nonumber \\
& &ik( e^{-ika/2}-R(k) e^{ika/2})=i \hat k( A_{+}(k) e^{-i{\hat k}a/2}-A_{-}(k) e^{i{\hat k}a/2}),  
\end{eqnarray}
and at $x=a/2$
\begin{eqnarray}
& &  A_{+}(k) e^{i{\hat k}a/2}+A_{-}(k) e^{-i{\hat k}a/2}= T(k) e^{ika/2},   \nonumber \\ 
& & i \hat k(  A_{+}(k) e^{i{\hat k}a/2}-A_{-}(k) e^{-i{\hat k}a/2})=  ik  T(k) e^{ika/2}.
\end{eqnarray}
Solving these equations, we have the following solutions: 
\begin{eqnarray}
R(k)&=&  e^{-ika} \frac{(k^2-{\hat k}^2)\sin \hat k a}{D(k)}  ,~~T(k)= e^{-ika} \frac{2i k \hat k}{D(k)}, \\
A_{\pm}(k)&=&e^{-i (k \pm \hat k)a/2}(1 \pm \frac{k}{\hat k}) \frac{i k \ \hat k}{D(k)} , \\
D(k)&=&(k^2+{\hat  k}^2)\sin {\hat k}a +2ik \hat k \cos {\hat k}a. \label{square_solution}
\end{eqnarray}
The total flux  satisfies Eq.$(\ref{unit_flux})$,
$|T(k)|^2+|R(k)|^2=1$.
 Values at the boundary are given by
\begin{eqnarray}
\phi(k,-a/2)&=& e^{-ika/2}\frac{2k^2 \sin {\hat k a} +2i k \hat k \cos \hat k a }{D(k)},\\
\phi(k,a/2)&=&  e^{-ika/2}\frac{2i k \hat k}{D(k)}.
\end{eqnarray}
The ratio of the values at $x= \pm \frac{a}{2}$ is  
\begin{eqnarray}
\frac{|\phi(k,-a/2)|^2 }{|\phi(k,a/2)|^2}=\frac{k^2 \sin^2{\hat k}a +{\hat k}^2 \cos^2 {\hat k}a}{{\hat k}^2} \neq 1,
\end{eqnarray}
and their difference is 
\begin{eqnarray}
|\phi(k,-a/2)|^2-|\phi(k,a/2)|^2
&=&\frac{4k^2}{|D(k)|^2}(k^2-{\hat k}^2) \sin^2 {\hat k} a. \label{difference}
\end{eqnarray}
These are understandable as Eq.$(\ref{eigenstates_sp})$ is not symmetric under  $x \rightarrow -x$.  
\subsection{Scalar product of stationary states :direct integrals}
We first compute the scalar product of $ \phi(k_2,x)$ and $ \phi(k_1,x)$, 
\begin{eqnarray}
& &I_0=\int_{-\infty}^{\infty} dx( \phi({k_2},x)^{*}   \phi(k_1,x) ), 
\end{eqnarray}
directly.  The integral 
\begin{eqnarray}
& &I_0=\int_{-\infty}^{-a/2} dx  ( e^{-ik_2x}+R(k_2)^{*} e^{ik_2x})   (e^{ik_1x}+R(k_1) e^{-ik_1 x} )\nonumber \\
& &+ \int_{-a/2}^{a/2} dx  (  A_{+}(k_2)^{*} e^{-i{\hat k_2}x}+A_{-}(k_2)^{*} e^{i{\hat k_2}x})  ( A_{+}(k_1) e^{i{\hat k_1}x}+A_{-}(k_1) e^{-i{\hat k_1}x}) \nonumber\\
& &+ \int_{a/2}^{\infty} dx   T(k_2)^{*} e^{-ik_2x} T(k_1) e^{ik_1x}, 
\end{eqnarray}
is expressed using  an integral     
\begin{eqnarray}
I(X_1,X_2;k_1-k_2)=\int_{X_1}^{X_2} dx e^{i(k_1-k_2)x}, \label{integral}
\end{eqnarray}
as
\begin{eqnarray}
& &I_0=I(-\infty,-a/2;k_1-k_2)+I(-\infty,-a/2;-k_1-k_2)R(k_1) \nonumber \\
& &+ I(-\infty,-a/2;k_1+k_2)R(k_2)^{*}+I(-\infty,-a/2;k_2-k_1) R(k_2)^{*} R(k_1) \nonumber \\
& &+A_{+}(k_2)^{*}A_{+}(k_1) I(-a/2,a/2; \hat k_1-\hat k_2)+A_{+}(k_2)^{*} A_{-}(k_1)I(-a/2,a/2; -\hat k_2-\hat k_1) \nonumber \\
& &+A_{-}(k_2)^{*} A_{+}(k_1) I(-a/2,a/2; \hat k_2+\hat k_1)+A_{-}(k_2)^{*} A_{-}(k_1)I(-a/2,a/2; \hat k_2-\hat k_1 )\nonumber \\
& &+T(k_2)^{*} T(k_1) I(a/2, \infty; -k_2+k_1). 
\end{eqnarray}

%
%
\subsubsection{Reduced integrals}

 Applying Eqs.$(\ref{stationary_2})$-$(\ref{stationary_4})$,  $(\ref{comparison_integrals_1})$, and $(\ref{comparison_integrals_2})$ for a wave function
\begin{eqnarray}
\phi(k,x)=e^{ikx}+R(k) e^{-ikx}, x<-a/2 ; ~~T(k) e^{ikx} x>a/2, 
\end{eqnarray}
we compute $J(k_2,k_1)_{x_1}^{x_2} $ for  $x_1<-a/2, x_2>a/2$ ,  $x_1<-a/2, x_2<-a/2$, and $x_1<-a/2, x_2<-a/2$. This is expressed for  $x_1<-a/2, x_2>a/2$ as,
\begin{eqnarray}
& &J(k_2,k_1)_{x_1}^{x_2} \nonumber \\
& &=-i(k_1+k_2)( (T(k_2)^{*} T(k_1) e^{i (k_1-k_2) x_2}- e^{i (k_1-k_2) x_1})+e^{-i(k_1-k_2)x_1} R(k_2)^{*}R(k_1)) \nonumber \\
& &+i(k_1-k_2) (R(k_1)e^{-i(k_1+k_2)x_1}- R(k_2)^{*} e^{i(k_1+k_2)x_1}),
\end{eqnarray}
and for $x_1<-a/2, x_2<-a/2$,    
\begin{eqnarray}
& &J(k_2,k_1)_{x_1}^{x_2} \nonumber \\
& &=-i(k_1+k_2)( (e^{i(k_1-k_2)x_2}- e^{i (k_1-k_2) x_1})+(e^{-i(k_1-k_2)x_1}-e^{i(k_1-k_2)x_2}) R(k_2)^{*}R(k_1)) \nonumber \\
& &-i(k_1-k_2) (R(k_1)(e^{-i(k_1+k_2)x_1}-e^{-i(k_1+k_2)x_2})- R(k_2)^{*} (e^{i(k_1+k_2)x_1}-e^{i(k_1+k_2)x_2}), \nonumber \\
& &
\end{eqnarray}
and for $x_1>a/2, x_2>a/2$,   
\begin{eqnarray}
& &J(k_2,k_1)_{x_1}^{x_2} =-i(k_1+k_2)( (e^{i(k_1-k_2)x_2}- e^{i (k_1-k_2) x_1})T(k_2)^{*}T(k_1)) ~~
\end{eqnarray}

We evaluate  $J(k_2,k_1)_{x_1}^{x_2} $ at $x_2 \rightarrow \infty$ and $x_1 \rightarrow -\infty$ using a formula of a free wave Eq.$(\ref{delta_m})$ and the unitarity  Eq.$(\ref{unit_flux})$. 
Then we have
\begin{eqnarray}
& &-\frac{1}{2(E_2-E_1)}J(k_2,k_1)_{x_1}^{x_2} \nonumber \\
%
%
& &=  2  \pi \delta(k_1-k_2)+(R(k_1)+R(k_2)^{*}) \pi \delta(k_1+k_2) +\Delta, \label{scalar-product_well}
\end{eqnarray}
where
\begin{eqnarray}
\Delta=( T(k_2)^{*}T(k_1)-1+R(k_2)^{*}R(k_1))\frac{1}{i(k_1-k_2)}+(R(k_1)-R(k_2)^{*} )\frac{1}{i(k_1+k_2)}. \label{non-conserving_well} \nonumber \\
& &
\end{eqnarray}
These are equivalent to  Eqs.$(\ref{scalar_product})$ and $(\ref{non-conserving})$. By separating  $J(k_2,k_1)_{x_1}^{x_2} $ into three components, we have
\begin{eqnarray}
J(k_2,k_1)_{x_1}^{x_2}=J(k_2,k_1)_{x_1}^{-a/2}+J(k_2,k_1)_{-a/2}^{a/2}+J(k_2,k_1)_{a/2}^{x_2}.
\end{eqnarray}
Each term in the right-hand side is  expressed by the values at boundaries.  We have Eq.$(\ref{scalar-product_well})$ for sum of three terms.  

\begin{figure}[t]
\includegraphics[width=1.0\textwidth]{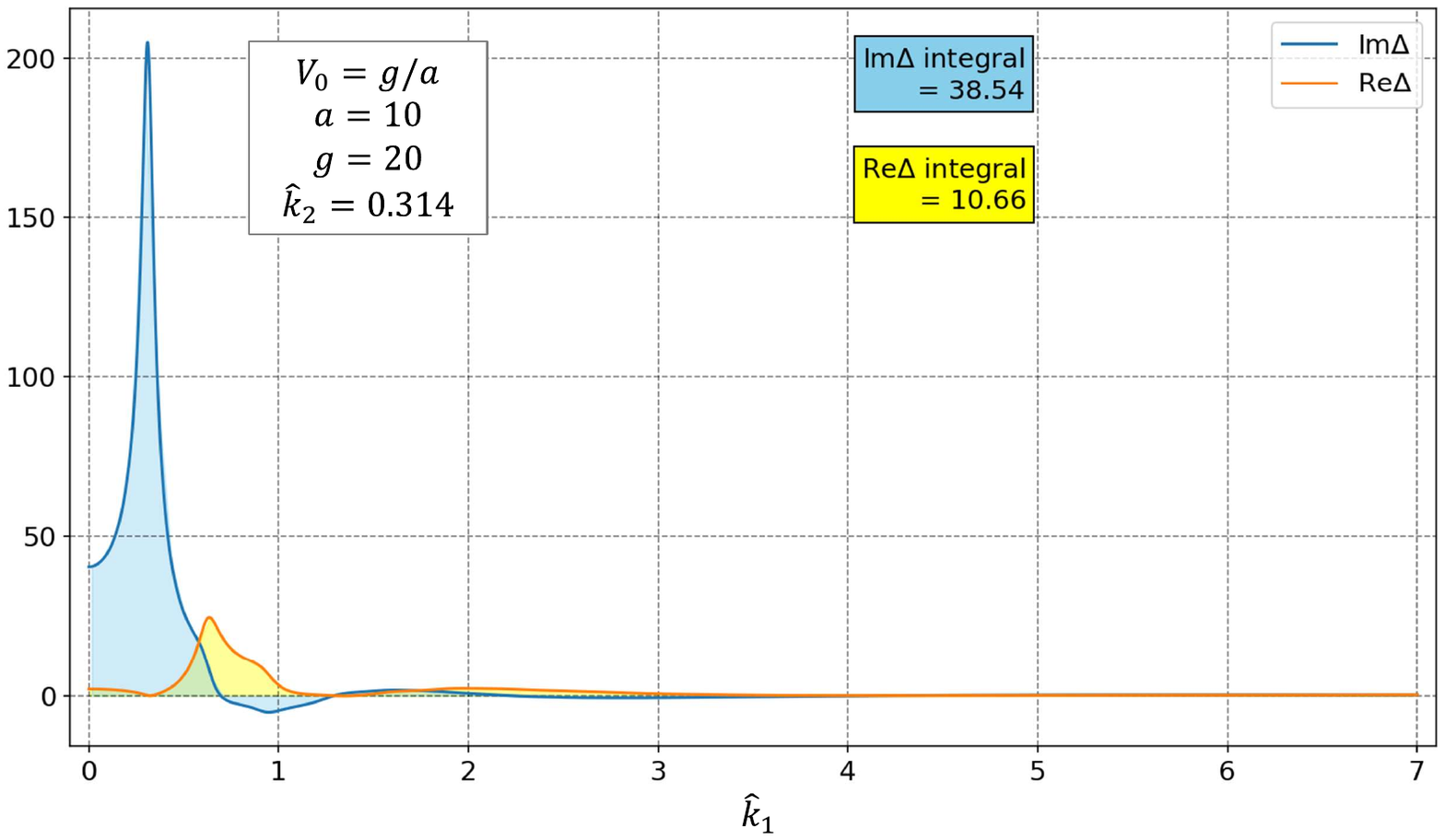}
\caption{  Imaginary and real parts of $\Delta $ at $a \hat k_2=0.314 $ is plotted as a function of $\hat k_1$ for $a=10, g=20$ in  $\hbar=1$ and $m=1$.   The imaginary part of $\Delta $ is maximum at $ k_1=0.314 $ and decreases rapidly  at larger or smaller $ k_1$ and the real part has a peak at a larger $k_1$ Areas of the imaginary and real parts are 38.54 and 10.66.  }
\label{fig:square_well_2}
\end{figure}

Substituting Eq.$(\ref{square_solution})$  into $R(k)$ and $T(k)$, we have   $\Delta \neq 0 $.  Figures. 1-4  show   $ \Delta $ and $|\Delta|^2$. Fig.1 shows the real 
 and imaginary parts of $\Delta$ for the potential $V_0=2, a=10, g=20$ at $\hat k_2=0.314$. Note that $\hbar=1$ and $m=1$ are used. The area of the imaginary part is $38.54$ and the area of the real part is $10.66$ 
whereas the area of the momentum conserving term is $2\pi$ from Eq.$(\ref{scalar-product_well})$. 
\begin{figure}[t]
\includegraphics[width=1.0\textwidth]{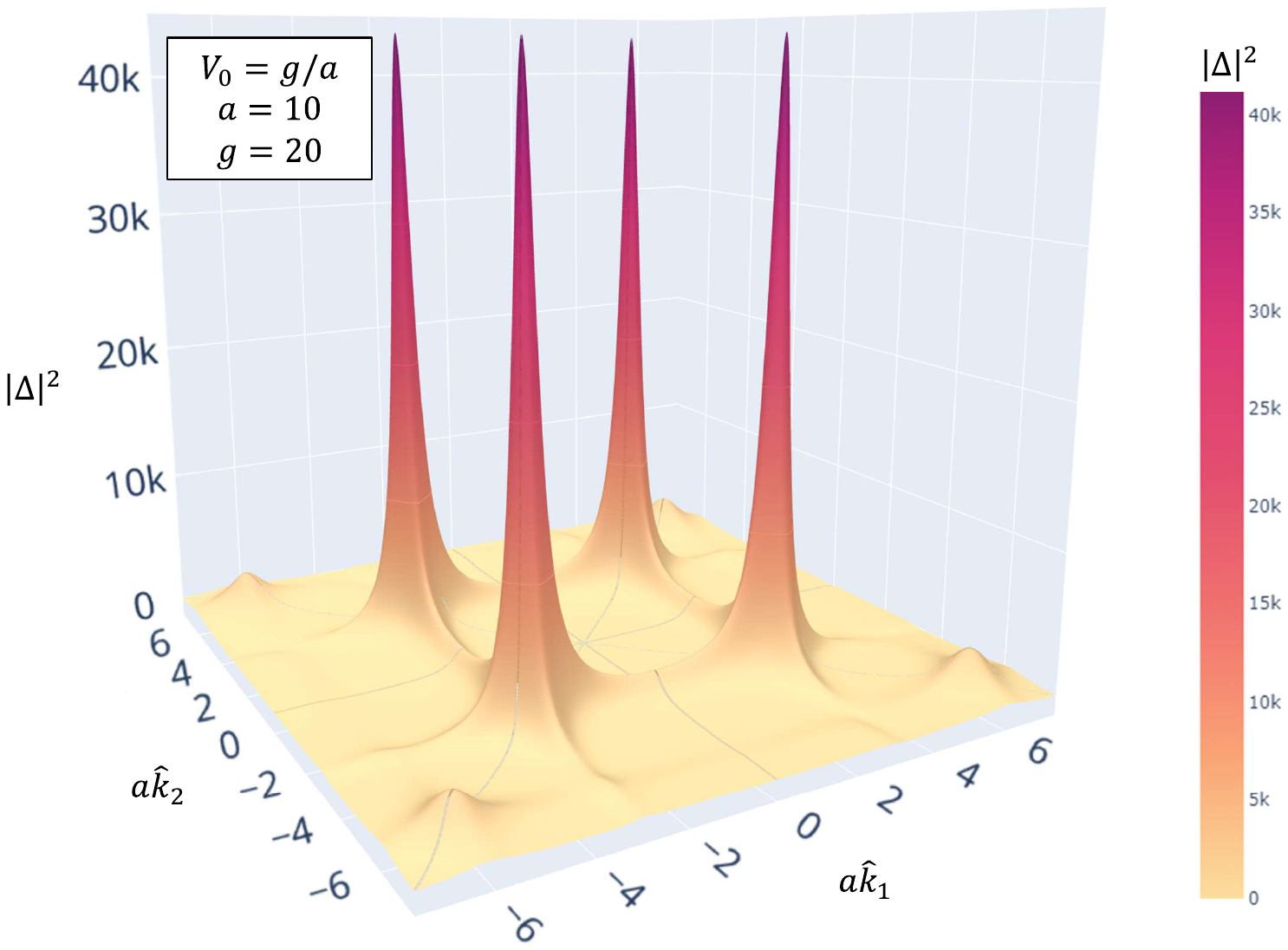}
\caption{  $|\Delta |^2$  is plotted in the vertical direction for parameters $a=10, g=20$ in  $\hbar=1$ and $m=1$. Holizontal directions stand for $a\hat k_1$ and $a \hat k_2$. $|\Delta |^2$ has peaks at $ a \hat k_1= a \hat k_2= \pm \pi$ and decreases fast with   $ |\hat k_1 -   \hat k_2|$.}
\label{fig:square_well}
\end{figure}

 We evaluate at   momenta  
\begin{eqnarray}
\hat k_i  a= n_i \pi, n_i=0,1,2, \cdots.
\end{eqnarray}
It follows that 
\begin{eqnarray}
R(k)&=&  e^{-ika} \frac{(k^2-{\hat k}^2)\sin \hat k a}{D(k)} =0 ,\\
T(k)&=& e^{-ika} \frac{2i k \hat k}{D(k)}=(-1)^{n_i} e^{-ika}, \\
D(k)&=&(k^2+{\hat  k}^2)\sin {\hat k}a +2ik \hat k \cos {\hat k}a=2ik \hat k (-1)^{n_i},
\end{eqnarray}
and 
\begin{eqnarray}
& &-\frac{1}{2(E_2-E_1)}J(k_2,k_1)_{x_1}^{x_2}= 2\pi \delta(k_1-k_2) +i((-1)^{n_2-n_1}e^{i(k_1-k_2)a}-1)\frac{1}{k_1-k_2}. \nonumber \\
& & 
\end{eqnarray}
Non-diagonal terms remain. In Fig.2, $|\Delta|^2$ is plotted, and .in Fig. 3 and Fig.4 , the imaginary and real parts of $ \Delta$are plotted.  
\begin{figure}[t]
\centering
\includegraphics[width=1.0\textwidth]{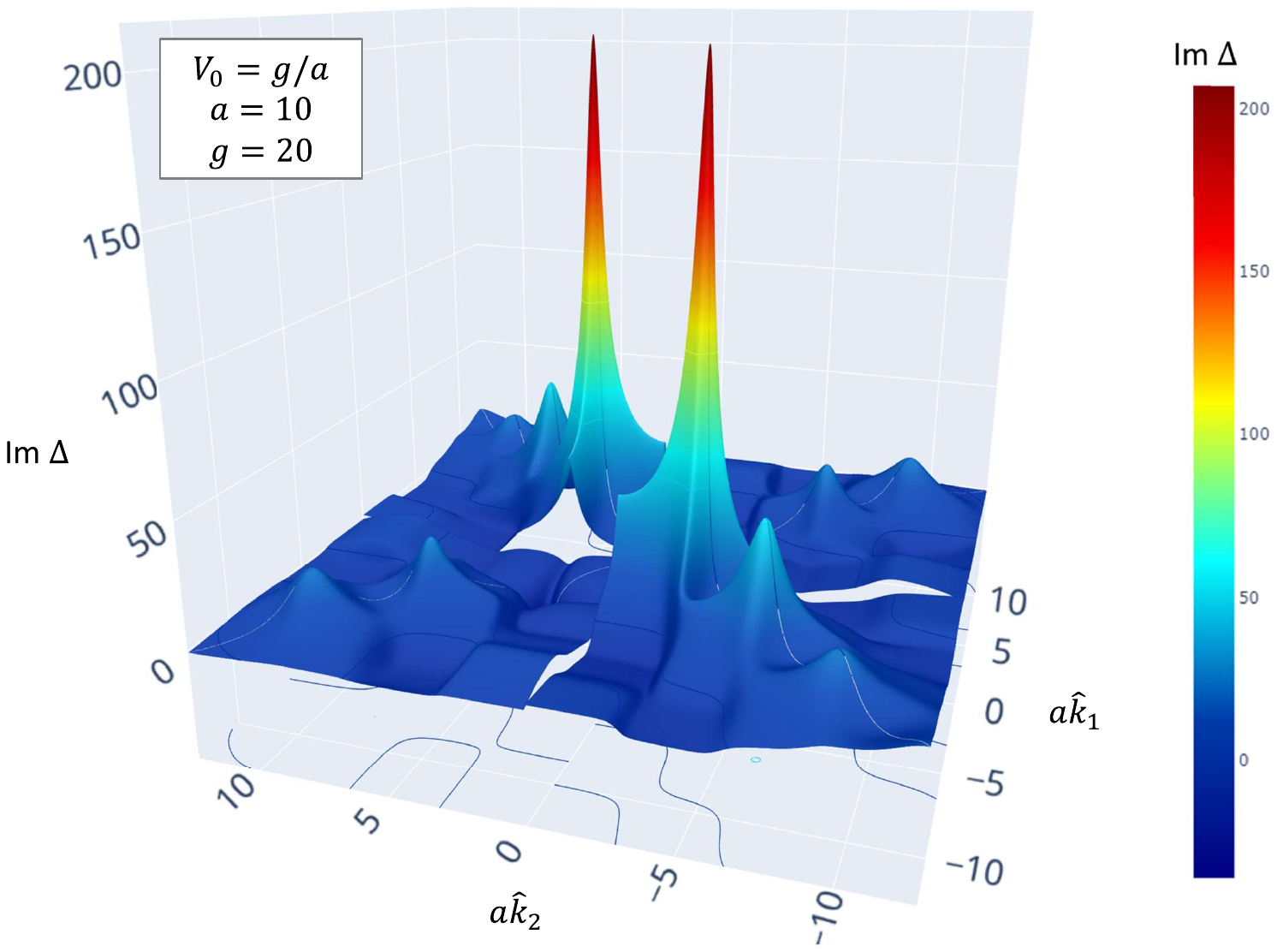}
\caption{ Imaginary part of  $\Delta $  is plotted in the vertical direction for parameters $a=10, g=20$ in  $\hbar=1$ and $m=1$.  Holizontal directions stand for $a\hat k_1$ and $a \hat k_2$. Imaginary part of $\Delta $ has peaks at $ a \hat k_1= a \hat k_2= \pm \pi$ and decreases fast with   $ |\hat k_1 -   \hat k_2|$.}
\label{fig:square_well_3}
\end{figure}
\begin{figure}[t]
\includegraphics[width=1.0\textwidth]{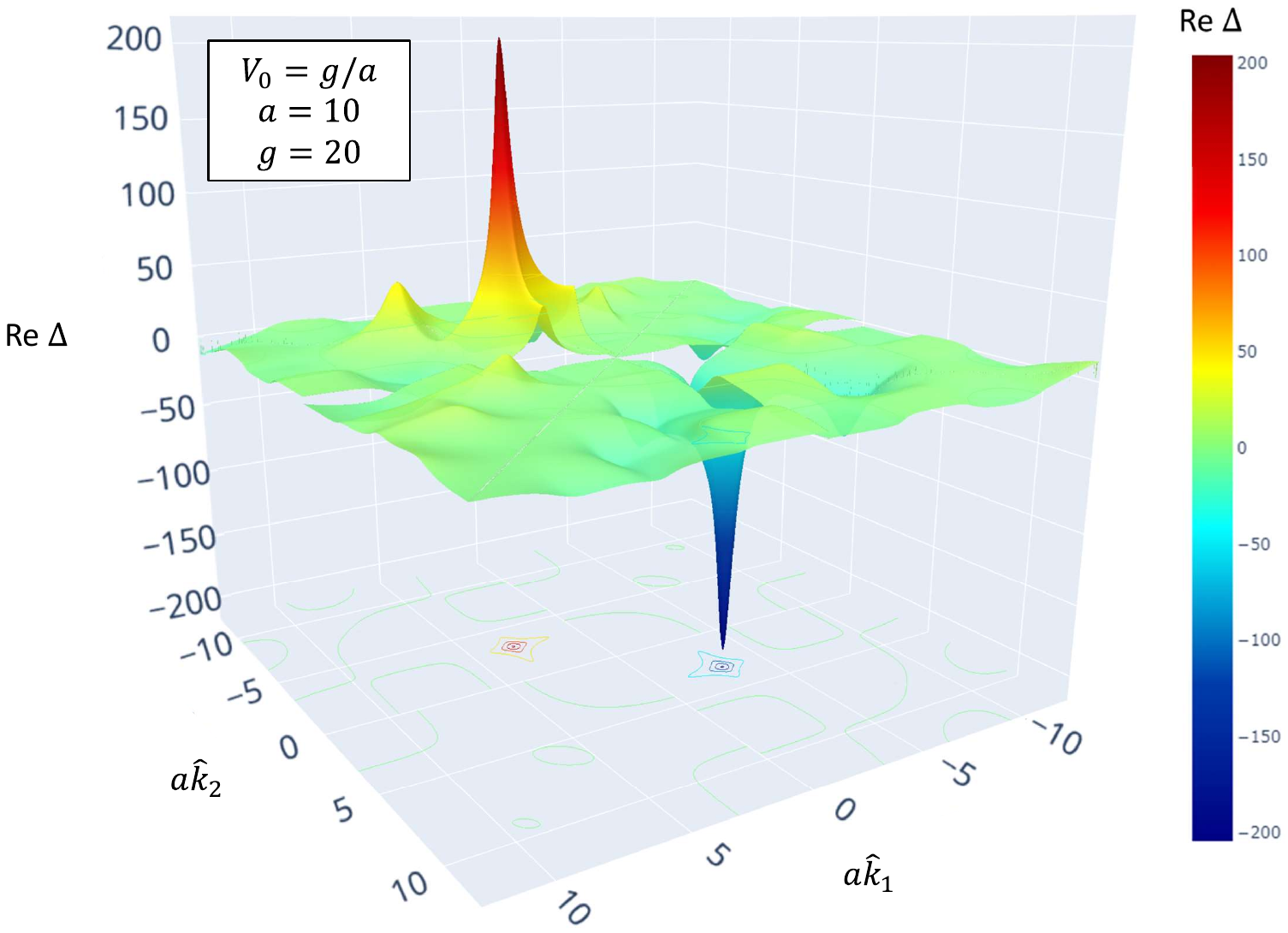}
\caption{ Real part of  $\Delta  $  is plotted in the vertical direction for parameters $a=10, g=20$ in  $\hbar=1$ and $m=1$.  Holizontal directions stand for $a\hat k_1$ and $a \hat k_2$. Real  part $\Delta $ has peaks at $ a \hat k_1= a \hat k_2= \pm \pi$ and decreases fast with   $ |\hat k_1 -   \hat k_2|$.}
\label{fig:square_well}
\end{figure}
\subsubsection{Regularized integrals}
 Scalar products of scattering stationary states with a regularized integral 
\begin{eqnarray}
\int_0^{\infty} dx e^{i(p+i\epsilon) x}=\frac{-1}{ i(p+i\epsilon)}, \label{regularized}
\end{eqnarray}
are studied in this subsection and Appendices. The integrals Eq.$(\ref{integral})$ are given  as below, 
\begin{eqnarray}
& &I^{\epsilon}(-\infty,-a/2;k_1-k_2)=e^{-i(k_1-k_2) a/2} \frac{1}{i}( P\frac{1}{k_1-k_2} +i \pi \delta(k_1-k_2)),  \nonumber \\
& & \\
& &I^{\epsilon}(a/2,\infty;k_1-k_2)= e^{i(k_1-k_2)a /2} \frac{1}{i}(- P\frac{1}{k_1-k_2} +i \pi \delta(k_1-k_2)), \nonumber \\
& &\\
& &I^{\epsilon}(-a/2,a/2;\hat k_1-\hat k_2)=\frac{1}{i(\hat k_1-\hat k_2)} (e^{i(\hat k_1-\hat k_2)a/2}-e^{-i(\hat k_1-\hat k_2)a/2}). \nonumber \\
\end{eqnarray}
Accordingly, we have the expressions
\begin{eqnarray}
I_0&= &I_0^1+I_0^2, \\
I_0^1&=&I^{\epsilon}(-\infty,-a/2;k_1-k_2)+I^{\epsilon}(-\infty,-a/2;k_2-k_1) R(k_2)^{*} R(k_1)\nonumber\\
& &+T(k_2)^{*} T(k_1) I^{\epsilon}(a/2, \infty; -k_2+k_1)  \nonumber\\
& &+A_{+}( k_2)^{*}A_{+}( k_1) I^{\epsilon}(-a/2,a/2; \hat k_1-\hat k_2)+A_{-}(k_2)^{*} A_{-}(k_1)I^{\epsilon}(-a/2,a/2; \hat k_2-\hat k_1 ), \nonumber \\
& &    \\
I_0^2&=&+I^{\epsilon}(-\infty,-a/2;-k_1-k_2)R(k)+ I^{\epsilon}(-\infty,-a/2;k_1+k_2)R(k_2)^{*} \nonumber \\
& &+A_{+}( k_2)^{*} A_{-}( k_1)I^{\epsilon}(-a/2,a/2; -\hat k_2-\hat k_1)+A_{-}( k_2)^{*} A_{+}( k_1) I^{\epsilon}(-a/2,a/2; \hat k_2+\hat k_1). \nonumber\\
& & 
\end{eqnarray}

It follows that
\begin{eqnarray}
I_0^1&=& \pi \delta(k_1-k_2) (1+|R(k_1)|^2+ |T(k_1)|^2 ) \nonumber \\
&+&  \frac{1}{i} P\frac{1}{k_1-k_2} (e^{-i(k_1-k_2) a/2} -e^{i(k_1-k_2) a/2} (R(k_2)^{*} R(k_1) +T(k_2)^{*} T(k_1)   )  
\nonumber \\
&+&   \frac{2\sin((\hat k_1-\hat k_2) a/2)}{\hat k_1-\hat k_2} ( A_{+}( k_2)^{*} A_{+}( k_1)  +A_{-}( k_2)^{*} A_{-}( k_1) ), 
\end{eqnarray}
where
\begin{eqnarray}
& & (e^{-i(k_1-k_2) a/2} -e^{i(k_1-k_2) a/2}     (R(k_2)^{*} R(k_1) +T(k_2)^{*} T(k_1))  ) \nonumber \\
& & = \frac{e^{-i(k_1-k_2) a/2}}{D(k_2)^{*} D(k_1)} (D(k_2)^{*}D(k_1)  -        {4V_0^2 \sin {{{\hat k_2}}} a} { \sin \hat k_1 a}  - { 2 k_2 \hat {k_2} } {( 2 k_1 {\hat k_1} )} ), 
\nonumber \\
\end{eqnarray}
and 
\begin{eqnarray}
& &A_{+}(k_2)^{*} A_{+}( k_1)  +A_{-}( k_2)^{*} A_{-}(k_1) =  \frac{ 1  }{D(k_2)^{*} D(k_1)} k_1 k_2 \nonumber \\
& & \times ( (k_2+\hat k_2)(k_1+ \hat k_1)e^{i(k_2+\hat k_2-k_1-\hat k_1)a/2}+(\hat k_2 -k_2)( \hat k_1-k_1)  e^{i(k_2- \hat k_2-k_1+ \hat k_1)a/2}). \nonumber \\
\end{eqnarray}
We have finally
\begin{eqnarray}
& &I_0^1=\pi \delta(k_1-k_2) (1+|R(k_1)|^2+ |T(k_1)|^2 ) \nonumber \\
& &+\frac{1}{D(k_2)^{*} D(k_1)} (\frac{1}{i} P {\frac{  e^{-i(k_1-k_2) a/2}} {k_1-k_2} }  (D(k_2)^{*}D(k_1)  -        {4V_0^2 \sin {{{\hat k_2}}} a} { \sin \hat k_1 a}  - { 2 k_2 \hat {k_2} } {( 2 k_1 {\hat k_1} )} )\nonumber \\        
& &+  \frac{2\sin((\hat k_1-\hat k_2) a/2)}{\hat k_1-\hat k_2} k_1 k_2  ( (k_2+\hat k_2)(k_1+ \hat k_1)e^{i(k_2+\hat k_2-k_1-\hat k_1)a/2} \nonumber \\
& &+(\hat k_2 -k_2)( \hat k_1-k_1)  e^{i(k_2- \hat k_2-k_1+ \hat k_1)a/2})) \label{scalar-product_1}
\end{eqnarray}
and
\begin{eqnarray}
& &I_0^2=\frac{1}{i}( P\frac{-1}{k_1+k_2} )  2V_0 \frac{e^{i(k_2-k_1)a/2}}{D(k_1)D(k_2)^{*}}[ 2(k_2^2-k_1^2) \sin {\hat k_1}a \sin {\hat k_2}a \nonumber \\
& & ~~-2ik_2 \hat k_2 \cos {\hat k_2}a \sin {\hat k_1}a -2ik_1 \hat k_1 \cos {\hat k_1}a \sin {\hat k_2}a ]  + (  \pi )2 e^{-ik_1a} 2V_0 \frac{\sin \hat k_1a}{D(k_1)} \delta(k_1+k_2) \nonumber \\
& &+     \frac{1}{D(k_2)^{*}D(k_1)} [  (k_2k_1)((\hat k_2+k_2)(\hat k_1-k_1)e^{i(k_2+\hat k_2 -k_1+ \hat k_1)a/2} \nonumber \\
& &+(\hat k_2-k_2)(\hat k_1+k_1) e^{i(k_2+\hat k_2-k_1-\hat k_1)a/2}) ] \frac{1}{i(\hat k_1+\hat k_2)}2i\sin(\hat k_1+\hat k_2)a/2). \nonumber \\
& &\label{scalar-product_2}
\end{eqnarray}
The first line of the inner product represented in Eqs.$(\ref{scalar-product_1})$ and $ (\ref{scalar-product_2}) $ is proportional to $\delta(k_1-k_2)$ but the other terms are not.  
 

Adding $I_0^1$ and $I_0^2$,  we have,
\begin{eqnarray}
& &I_0=2 \pi \delta(k_1-k_2)   +   \pi 2 e^{-ik_1a} 2V_0 \frac{\sin \hat k_1a}{D(k_1)} \delta(k_1+k_2) +\Delta \label{scalar-product_3}
\end{eqnarray}
where
\begin{eqnarray}
& &\Delta=\frac{1}{D(k_2)^{*} D(k_1)} (\frac{1}{i} P {\frac{  e^{-i(k_1-k_2) a/2}} {k_1-k_2} }  (D(k_2)^{*}D(k_1)  -        {4V_0^2 \sin {{{\hat k_2}}} a} {  \sin \hat k_1 a}  \nonumber \\
& & - { 2 k_2 \hat {k_2} } {( 2 k_1 {\hat k_1} )} )            +  \frac{2\sin((\hat k_1-\hat k_2) a/2)}{\hat k_1-\hat k_2} k_1 k_2 \nonumber\\
& &\times (  ( (k_2+\hat k_2)(k_1+ \hat k_1)e^{i(k_2+\hat k_2-k_1-\hat k_1)a/2}+(\hat k_2 -k_2)( \hat k_1-k_1)  e^{i(k_2- \hat k_2-k_1+ \hat k_1)a/2})      ) \nonumber \\
& &+\frac{1}{i}( P\frac{-1}{k_1+k_2} )  2V_0 \frac{e^{i(k_2-k_1)a/2}}{D(k_1)D(k_2)^{*}}[ 2(k_2^2-k_1^2) \sin {\hat k_1}a \sin {\hat k_2}a   \nonumber \\
& & ~~-2ik_2 \hat k_2 \cos {\hat k_2}a \sin {\hat k_1}a -2ik_1 \hat k_1 \cos {\hat k_1}a \sin {\hat k_2}a ]   \nonumber \\
& &+     \frac{k_2 k_1}{D(k_2)^{*}D(k_1)}((\hat k_2+k_2)(\hat k_1-k_1)e^{i(k_2+\hat k_2 -k_1+ \hat k_1)a/2}  +(\hat k_2-k_2)(\hat k_1+k_1) e^{i(k_2+\hat k_2-k_1-\hat k_1)a/2}) \nonumber \\
& & \frac{1}{i(\hat k_1+\hat k_2)} 2i\sin(\hat k_1+\hat k_2)a/2).\label{scalar-product_4}
\end{eqnarray}
From Eqs.$( \ref{scalar-product_3})$ and $( \ref{scalar-product_4})$,  it is observed that the two stationary states with  different energies are not orthogonal. The non-orthogonality of the regularized integrals  is consistent with the non-orthogonality of Eq.$(\ref{non-conserving})$.

\subsection{ A solvable potential   }

 For a potential $V(x)=V_0\frac{1}{\cosh^2 \mu x} $, and $m=1$ and $\mu=1$
\begin{eqnarray}
& &R(k)= \frac{\Gamma(ik)\Gamma(1+\nu-ik)\Gamma(-\nu-ik)}{\Gamma(-ik)\Gamma(1+\nu)\Gamma(-\nu)},   \\
& &T(k)= \frac{\Gamma(1+\nu-ik)\Gamma(-\nu-ik)}{\Gamma(-ik)\Gamma(1-ik)}. 
\end{eqnarray}
\cite{Landau-Lifshits}.The non-conserving term is given by Eq.$(\ref{non-conserving})$.

 In these short-range potentials, the nonconserving   term $\Delta$ do not vanish, and the stationary scattering states of  different energies are not orthogonal. These results are  consistent with \cite{Landau-Lifshits}.  

The Non-conserving term  $\Delta$   in the potential $V_0 \frac{1}{\cosh^2 \mu x}$  was obtained from asymptotic behavior, which  has a deviation that disappears in a limit  $x \rightarrow \infty$. The   small deviation of the wave function from the asymptotic value $\delta \phi=\phi-\phi_{asy}$ leads to a small correction to $\Delta$. Its magnitude is estimated.      
\begin{eqnarray}
& &\delta J(k_2,k_1)_{x_1}^{x_2}=\phi(k_2,x_2)^{*}(-\frac{d}{dx})\delta \phi_1(x_2)+\delta \phi_2^{*}(x_2) (-\frac{d}{dx}) \phi(k_1,x_2)\nonumber \\
& &-\phi(k_1,x_2)(-\frac{d}{dx})\delta \phi_2^{*}(x_2)-\delta \phi_1(x_2) (-\frac{d}{dx}) \phi(k_2,x_2)^{*}-\phi(k_2,x_1)^{*}(-\frac{d}{dx})\delta \phi_1(x_1) \nonumber \\
& &+\delta \phi_2^{*}(x_1) (-\frac{d}{dx}) \phi(k_1,x_1)
+\phi(k_1,x_1)(-\frac{d}{dx})\delta \phi_2^{*}(x_1)-\delta \phi_1(x_1) (-\frac{d}{dx}) \phi(k_2,x_2)^{*}\nonumber\\
& &=O(e^{-\mu |x_i|}).
\end{eqnarray}
The correction is negligibly small for $|x_i| \gg \frac{1}{\mu}$.

Wave functions of bound states decrease rapidly with $x$. Accordingly,  $J(E_1,E_2)_{x_1}^{x_2} $ with  bound states of different energies vanish at $x_1 \rightarrow -\infty, x_2 \rightarrow \infty$. Products  of wave functions of  bound and scattering states decrease rapidly and these states  are orthogonal. 
\subsection{Spherically symmetric potential in three  dimensions }
Similarly, an equation for  a radial variable in a spherically symmetric potential in three dimensions is studied.  We use a radial variable $r \geq 0$ in a potential $V(r)$ for  $\hbar=1$ and $m=1$. Eigenvalue equation for a spherically symmetric solution  is,
\begin{eqnarray}
& &(-\frac{1}{2}\frac{d^2}{dr^2}+V(r))\phi(k,r)=E(k) \phi(k,r),\\
& &(-\frac{1}{2}\frac{d^2}{dr^2}+V(r))\phi(k,r)^{*}=E(k)^{*} \phi(k,r)^{*}. \nonumber
\end{eqnarray}
and has asymptotic behaviors 
\begin{eqnarray}
\phi(r)&=&T(k)e^{ikr}+R(k) e^{-ikr}, r_0<r. \label{asymptotoic_2}
\end{eqnarray}
Multiply the complex conjugates, and relations equivalent to  Eqs.$(\ref{comparison_integrals_1})$ and $( \ref{comparison_integrals_2})$  
\begin{eqnarray}
& &\int_{0}^{r_2}dr \frac{d}{dr}\left( \phi(k_2,r)^{*}(-\frac{d}{dr})\phi(k_1,r)-\phi(k_1,r)(-\frac{d}{dr})\phi(k_2,r)^{*}\right)\nonumber\\
& &~~=(E(k_1)-E(k_2)^{*}) \int_{0}^{r_2} dr \phi(k_2,r)^{*}\phi(k_1,r).
\end{eqnarray}
is derived.The left-hand side is the difference of the current at two boundaries,  
\begin{eqnarray}
& &J(k_2,k_1)_{0}^{r_2}\nonumber\\
& &~~=-i(k_1+k_2) T(k_2)^{*}T(k_1)e^{-i(k_2-k_1)r_2}+i(k_1+k_2) R(k_2)^{*}R(k_1)e^{i(k_2-k_1)r_2}\nonumber \\
& &~~+T(k_2)^{*}R(k_1)e^{-i(k_1+k_2)r_2} i(k_1-k_2)-R(k_2)^{*} T(k_1)e^{i(k_1+k_2)r_2}i(k_1-k_2) \nonumber \\
& & ~~-( \phi(k_2,0)^{*}(-\frac{d}{dr})\phi(k_1,0)-\phi(k_1,0)(-\frac{d}{dr})\phi(k_2,0)^{*}),
\end{eqnarray}
which is reduced to 
\begin{eqnarray}
& &~~-i(k_1+k_2) ( T(k_2)^{*}T(k_1)e^{-i(k_2-k_1)r_2}- R(k_2)^{*}R(k_1)e^{i(k_2-k_1)r_2}) \nonumber \\
& &~~+ i(k_1-k_2)( T(k_2)^{*}R(k_1)e^{-i(k_1+k_2)r_2} -R(k_2)^{*} T(k_1)e^{i(k_1+k_2)r_2}) \nonumber \\
& & ~~-( \phi(k_2,0)^{*}(-\frac{d}{dr})\phi(k_1,0)-\phi(k_1,0)(-\frac{d}{dr})\phi(k_2,0)^{*}).
\end{eqnarray}

We evaluate this for $r_2 \rightarrow \infty$ using  formulae Eq.$(\ref{delta_m})$,
as 
\begin{eqnarray}
& &\frac{1}{k_1^2-k_2^2}J(k_2,k_1)_{0}^{\infty}\nonumber\\
& &~~=( T(k_2)^{*}T(k_1)+ R(k_2)^{*}R(k_1)) \pi \delta(k_1-k_2)  \nonumber \\
& &~~ ( T(k_2)^{*}R(k_1) +R(k_2)^{*} T(k_1)) \pi \delta(k_1+k_2)  +\Delta,\label{scalar-product_three}
\end{eqnarray}
where  nonorthogonal term, $\Delta$, is given by
\begin{eqnarray}
& &\Delta=-i\frac{1}{(k_1-k_2)} ( T(k_2)^{*}T(k_1)- R(k_2)^{*}R(k_1)) +i\frac{1}{(k_1+k_2)} (T(k_2)^{*}R(k_1) -R(k_2)^{*} T(k_1))\nonumber \\
& & ~~-[( \phi(k_2,0)^{*}(-\frac{d}{dr})\phi(k_1,0)-\phi(k_1,0)(-\frac{d}{dr})\phi(k_2,0)^{*})]\frac{1}{k_1^2-k_2^2}. \label{non-conserving_three}
\end{eqnarray}
The first line in $\Delta$ is a contribution from the wave function in asymptotic region and the second line is a contribution from that at the origin. From a boundary condition 
\begin{eqnarray}
(-\frac{d}{dr})\phi(k,0)=0
\end{eqnarray}
and 
\begin{eqnarray}
& &\Delta=-i\frac{1}{(k_1-k_2)} ( T(k_2)^{*}T(k_1)- R(k_2)^{*}R(k_1)) \nonumber \\
& &+i\frac{1}{(k_1+k_2)} (T(k_2)^{*}R(k_1) -R(k_2)^{*} T(k_1)). 
\end{eqnarray}

For a free wave,
\begin{eqnarray}
 & & T(k)=R(k)=1, \nonumber \\
& &\phi(k,0)=2,\frac{d}{dr} \phi(k,0)=0,
\end{eqnarray}
and 
\begin{eqnarray}
& &\Delta=0.
\end{eqnarray}
For stationary states in a short range potential, $\Delta$ does not vanish. In general, the stationary scattering states of  different energies are not orthogonal as well also in three dimensions,  which is  consistent with \cite{Landau-Lifshits}.

 Stationary solutions   of  continuous spectra of  Schr\"{o}dinger equations are studied in Section three and four.   Their scalar products are expressed using   the Dirac delta function for  plane-wave solutions and for various exceptional potentials. In these systems, the stationary states with the different energies are  orthogonal.  However, in   short-range potentials, the exact  solutions  of different energies are not orthogonal.  
Equations $(\ref{scalar_product})$ and $(\ref{non-conserving})$ are expressed by the reflection and transmission  coefficients, $R(k)$ and $T(k)$.
A  superposition of these waves does not have a  constant norm; therefore,  has a probability current  at  the infinity.

Non-orthogonality also holds   in other regularizations by taking  a finite upper bound $L$  and a limit $L \rightarrow \infty$ as shown in Eq.$(\ref{dirac_delta})$. The fact that stationary states with  different energies are not orthogonal has been  pointed out  in Landau-Lifshitz \cite{Landau-Lifshits}. We are not aware of  a regularization method that overcomes this property.

A  superposition of these waves does not have a  constant norm; therefore,  is  unsuitable for expressing  isolate physical states.

  \section{ Superposition of continuum stationary states  }
In this section,    the norm  of  superpositions of stationary continuum  states are computed.  Using them, we  clarify if these states  represent isolate states. 

The probability current of the state  expressed by Eq.$(\ref{current})$ 
is a physical quantity  at a position $x$. A  net current vanishes and a norm of  wave function is constant at an arbitrary time in isolated states.  To study the transition probability of isolate state,   the state must satisfy this condition. 
The superpositions of orthogonal  stationary  states with  different energies have time-independent norm and  vanishing net probability current, whereas those of non-orthogonal states do not hold these properties.  
\subsection{Orthogonal stationary states}
 Stationary states with different energies  studied in Section 3 are orthogonal, and superposition 
\begin{eqnarray}
\psi(t,x)=\int dk a(k) e^{\frac{E(k)}{i }t} \phi(k,x)  (\ref{superposition_s})\nonumber 
\end{eqnarray}
with a weight  $a(k)$ has  time-independent norm   
\begin{eqnarray}
N_0=\int dk_1  |a(k_1)|^2. \label{constant_norm}
\end{eqnarray}
Superpositions of stationary states of a free Hamiltonian, linear potential, uniform magnetic field, and periodic potentials have time-independent norms and represent isolated states. 
\subsection{Non-orthogonal stationary states: one dimension}
 Superpositions of non-orthogonal continuum states  Eq.$(\ref{superposition_s})$ 
studied in Section 4 are studied hereafter. We focus one dimensional system first.    The norm of this state Eq.$(\ref{norm_scattering_s})$
is expressed by function,  Eqs. $(\ref{scalar_product})$ and $(\ref{non-conserving})$, and by a value at the asymptotic spatial regions, $R(k)$ and $T(k)$. 
The norm and probability current  of this state is  expressed by Eq.$(\ref{current})$. The norm is expressed as follows:  
\begin{eqnarray}
& &N(t)=\int dx \psi(t,x)^{*}\psi(t,x) =N_0-\int dk_1 dk_2 a(k_2)^{*}a(k_1)e^{-\frac{E(k_2)-E(k_1)}{i }t} \times \nonumber \\
& &\left((i( T(k_2)^{*}T(k_1)-1+R(k_2)^{*}R(k_1))\frac{1}{k_2-k_1} +i(R(k_1)-R(k_2)^*)\frac{1}{k_1+k_2} \right) \nonumber \\
\end{eqnarray}
where $N_0$ is expressed by Eq.$(\ref{constant_norm} )$.
We have  a formula for a time-derivative of the norm
\begin{eqnarray}
& &\frac{d}{dt} N(t) = \nonumber \\
& &-\frac{1}{  2m } \left( (k \tilde T(t))^{*} \tilde T(t)-( k \tilde I(t))^{*} \tilde I(t)+(k \tilde R(t))^{*}\tilde R(t) +(k \tilde I(t) )^{*}  \tilde R(t)-(k\tilde R(t))^*\tilde I(t)\right ) \nonumber \\
& &-\frac{1}{  2m }\left(( \tilde T(t)^{*}(k \tilde T(t))-( \tilde I(t))^{*})( k \tilde I(t))+\tilde R(t)^{*}(k \tilde R(t))  - (\tilde I(t))^{*}(k \tilde R(t))+ \tilde R(t)^*  k \tilde I(t) \right), \label{net_current} \nonumber \\
& &
\end{eqnarray}
where  
\begin{eqnarray}
& &\int dk_1 a(k_1)e^{\frac{E(k_1)}{i }t} = \tilde I(t), ~\int dk_1 a(k_1)e^{\frac{E(k_1)}{i }t} k_1  = (k \tilde I(t)),\nonumber \\
& &\int dk_1 a(k_1)e^{\frac{E(k_1)}{i }t}T(k_1) = \tilde T(t), ~\int dk_1 a(k_1)e^{\frac{E(k_1)}{i }t}T(k_1)k_1 = (k \tilde T(t)), \nonumber \\
& &\int dk_1 a(k_1)e^{\frac{E(k_1)}{i }t}R(k_1) = \tilde R(t), ~\int dk_1 a(k_1)e^{\frac{E(k_1)}{i }t}R(k_1)k_1 = (k \tilde R(t)). \nonumber \\
& & \label{correlation_amplitude}
\end{eqnarray}
 Eq.$(\ref{net_current}  )$ is equivalent to  a net  probability current to the system, Eq.$(\ref{conservation})$.
 These waves do not have   constant norms; therefore,  are  unsuitable for expressing  isolate physical states. 

Next we  compute a time-derivative of the norm with a Gaussian wave packet of a central momentum$P_0$ and position $X_0$ at $t=0$. This is  expressed by 
\begin{eqnarray}
a(k)=Ne^{-\frac{1}{2 \sigma}(k-P_0)^2-i k X_0},N= (\sigma \pi)^{-1/4} ,
\end{eqnarray}
where $\sigma$ is a constant. 
Then the superposition
\begin{eqnarray}
& &\psi(0,x)=\int dk a(k)  \phi(k,x) ~~~~~~~~~~ (\ref{superposition_s}) \nonumber
\end{eqnarray} 
decreases rapidly with $|x|$. The wave function at $ x \gg a$  is expressed from Eq.$(\ref{superposition_s2} )$ as 
\begin{eqnarray}
& &\psi(t,x) \approx N(2 \sigma \pi)^{1/2}   e^{ iP_0 (x-X_0)-iE(P_0) t }  e^{-\frac{\sigma }{2}  (x-X_0-v_0 t)^2} T(k_0^{+})  \nonumber \\
\end{eqnarray}
and the one  at $ x \ll - a$  is expressed with stationary phase approximation  as 
\begin{eqnarray}
& &\psi(t,x) \approx N(2 \sigma \pi)^{1/2} (  e^{ iP_0 (x-X_0)-iE(P_0) t }  e^{-\frac{\sigma }{2}  (x-X_0-v_0 t)^2} \nonumber \\
& &+e^{ -iP_0 (x-X_0)-iE(P_0) t } R(k_0^{-}) e^{-\frac{\sigma }{2}  (x+X_0+v_0 t)^2} ),  \\
& &k_0^{\pm}= P_0-i \sigma(X_0\pm x+v_0t), v_0=\frac{P_0}{m},
\end{eqnarray} 
using a stationary phase approximation, where $a$ is a potential width and  $v_0$ is the group velocity of the wave packet.

Amplitudes provided by Eqs.$(\ref{correlation_amplitude})$  are computed with stationary phase approximations as
\begin{eqnarray}
& & \tilde I(t) = \psi_0(t) + \delta  \psi_0, \nonumber \\
 & &(   k  I(t))=  \psi_0(t)k_0(t)  + \delta ( k \hat I ),  \nonumber \\
& &\tilde T(t)=  \psi_0(t) T(k_0(t))  + \delta \tilde T(t),  \nonumber \\
& &(k \tilde  T(t))= \psi_0(t)k_0(t) T(k_0(t))  + \delta (k \tilde T(t)), 
\end{eqnarray} 
where 
\begin{eqnarray}
& &\psi_0(t)=e^{- iP_0 X_0-iE(P_0) t }N(2 \sigma \pi)^{1/2} e^{-\frac{\sigma }{2}  (X_0+v_0 t)^2}, \nonumber \\
& &k_0 (t)= P_0-i \sigma(X_0 +v_0t), 
\end{eqnarray}
and $ \delta  \psi_0$, $ \delta ( k \hat I )$,$ \delta \tilde T(t)$, and $ \delta (k \tilde T(t)) $,  are  next order terms.

Amplitudes of $R(k)$ are obtained by replacing $T(k)$ with $R(k)$. Ignoring small correction terms, we have 
\begin{eqnarray}
& &\frac{d}{dt}N(t) = -\frac{1}{  2m } |\psi_0(t)|^2 \times  \nonumber \\
& &
((k_0(t))^{*}\left( ( \tilde T(k_0(t))^{*} \tilde T(k_0(t))-1 + (\tilde R(k_0(t)))^{*}\tilde R(k_0(t)) + (  \tilde R(k_0(t))-(\tilde R(k_0(t)))^*\right ) \nonumber \\
& &+k_0(t) \left(( \tilde T(k_0(t))^{*}(\tilde T(k_0(t)))-1+\tilde R(k_0(t))^{*}( \tilde R(k_0(t)))  -(\tilde R(k_0(t)))+ \tilde R(k_0(t))^*   \right)). \nonumber \\
& &
\end{eqnarray} 

At  $t=T_0$ of satisfying $|X_0+v_0 T_0| =0$, $k_0=P_0$. Then
 $ \frac{d}{dt}N(t)=0$ , from unitarity relation, Eq.$(\ref{unit_flux})$. At $|t-T_0| \approx  0$,   $|\psi_0(t)|^2$ is sizable. $\frac{d}{dt} N(t)$ and flow of the probability current into the system becomes finite  when the wave packet is located in the potential region.  This  wave function has  the net flow of probability current. 

 Measurements in scattering processes are made when the wave packet is located far away from the potential region. Then  $|X_0+v_0 t| $ is large, and  $|\psi_0(t)|^2 = 0$. There is no current flow into the system in this period. Nevertheless, this state has   the current flow from outside  before the measurement, and is not  isolated. This condition has not been taken into account in the previous studies of the wave packet scattering in \cite{Schiff,Iwanami-kouza,Messia}. 

{\bf Lemma }

The condition for a state  been isolated  is an absence of  net current at any time.   Scattering matrix is applicable to scattering of isolate states only when the wave function satisfies the condition for  isolate states. 
 This condition  is necessary irrespect  of the  condition  at $t=\pm \infty$.
\subsection{Three dimensions}
Superposition of the continuum states, 
\begin{eqnarray}
\psi(t,\vec x)=\int d^3k a(\vec k) e^{\frac{E(k)}{i }t} \phi(\vec k, \vec x)  (\ref{superposition_s})\nonumber 
\end{eqnarray}    
of a weight  $a(\vec k)$.  The   norm
is expressed by wave function,  Eqs. $(\ref{scalar-product_three})$ and $(\ref{non-conserving_three})$, and behaves in the asymptotic spatial regions according to  coefficients,$R(k)$ and $T(k)$. 
The norm and probability current   is  expressed by Eq.$(\ref{current})$. The norm is expressed as follows: 
 Three dimensional system of a spherical potential is studied easily with radial variables $(r,\theta,\phi)$. The continuity relation reads 
\begin{eqnarray}
 & & \frac{\partial}{\partial t} \int_{r \leq R } d^3 x \rho(t,\vec x)+\int_{r=R} d {\vec  S} \cdot  { \vec j(t,\vec x)} =0 
\end{eqnarray}
where a volume integral is made over an inside of  sphere of radius $R$. 

Wave function of a stationary state, 
\begin{eqnarray}
\psi(x)=e^{ipz}+\frac{e^{ikr}}{r} f(\theta, \phi) \label{asymptotic_wave3} 
\end{eqnarray}
is used to compute a cross section, where the first term in the right-hand side represents an incoming  plane wave, and the second term represents a spherical wave. Because the first term in the right-hand side is expressed by a superposition of spherical waves   
 \begin{eqnarray}
e^{ipr \cos \theta}=\sum_l\frac{i}{2k}(2l+1) i^l P_l(\cos \theta)\left(\frac{e^{-i(kr-\frac{l \pi}{2})}}{r}-\frac{e^{i(kr-\frac{l \pi}{2})}}{r} \right)
\end{eqnarray}
the wave in the asymptotic region, Eq.$( \ref{asymptotic_wave3})$ is expressed as 
\begin{eqnarray}
\psi \approx \sum_l\frac{i}{2k}(2l+1) i^l P_l(\cos \theta)\left(\frac{e^{-i(kr-\frac{l \pi}{2})}}{r}-s_l(k)\frac{e^{i(kr-\frac{l \pi}{2})}}{r} \right),
\end{eqnarray}
where $s_l(k)$ is an amplitude for angular momentum $l$. For a s-wave, $l=0$, and   
\begin{eqnarray}
& &\psi_E \approx \frac{i}{2k}  \left(\frac{e^{-i(kr)}}{r}-s_0(k)\frac{e^{i(kr)}}{r} \right), r_0 <r  \label{s_wave}  \nonumber \\
& &\psi_E =\psi_E(b,x), r <r_0
\end{eqnarray}
This is a special case of Eq.$(\ref{asymptotoic_2})$  and is expressed as 
\begin{eqnarray}
 T(k)= -\frac{i}{2k} s_0(k),  R(k)=\frac{i}{2k}.
\end{eqnarray}
The norm and probability current  of this state is  expressed by Eq.$(\ref{current})$. The norm of superposition of spherical waves 
\begin{eqnarray}
\psi(r,t)= \int dk a(k) e^{\frac{E(k)}{i }t}\psi_E(r,t). \label{superposition_3}
\end{eqnarray}
is given by 
\begin{eqnarray}
& &N(t)=\int  r^2 dr d \Omega \psi(\vec x,t )^{*} \psi(\vec x,t )\nonumber \\
& &=4 \pi \int dr \phi(r)^{*}\phi(r) =N_0+4 \pi\int dk_1 dk_2 a(k_2)^{*}a(k_1)e^{-\frac{E(k_2)-E(k_1)}{i }t} \times \nonumber \\
& &i \left( \frac{1}{4k_1k_2}(s_0(k_2)^{*}s_0(k_1)-1) \frac{1}{k_2-k_1} +( \frac{s(k_1)-s_0(k_2)^*}{4k_1k_2}   )\frac{1}{k_1+k_2} \right) \nonumber \\
\end{eqnarray}
where 
\begin{eqnarray}
N_0=4 \pi\int dk_1  |a(k_1)|^2. 
\end{eqnarray}
From time-derivative, it follows that  
\begin{eqnarray}
& &\frac{d}{dt} N(t)
=\pi \int dk_1 dk_2 a(k_2)^{*}a(k_1) (-\frac{1}{ \hbar 2m }) e^{-\frac{E(k_2)-E(k_1)  }{i \hbar}t}    \frac{1}{k_1 k_2}\times \nonumber \\
& &\left((( s_0(k_2))^{*}( s_0(k_1)) -1)( k_1+k_2) +((- s_0(k_2))^{*}+  s_0(k_1))(k_2-k_1) \right) \nonumber \\
\end{eqnarray}

 For a Gaussian wave packet $a(k)=Ne^{-\frac{1}{2\sigma}(k-P_0)^2 -ikR_0}$, at time $t=T_0$ of satisfying $|R_0-v_0 T_0| \approx 0$,  $\tilde T(t)$ is sizable and at large $|R_0-v_0 t| $,  $\tilde T(t) = 0$. $\tilde R(t)$ is also the same. $\frac{d}{dt} N(t)$ and flow of the probability current into the system becomes finite  when the wave packet is located in the potential region.  At a later time when the wave packet is located away from  the potential region, there is no current flow into the system.  The current flow is expressed in the table.

Table of current fllow :
\\
\begin{tabular}{|c|c|c|c|}
\hline
  t   & $ \approx 0 $& $ \approx T_0$  & $ \infty$   \\
 \hline
$\frac{d}{dt}N(t) $     & 0 & $\neq 0  $   &$0$   \\
 \hline 
 \end{tabular}.
\\
\\
Wave function Eq.$(\ref{superposition_3})$ satisfies the condition for the isolated state in the asymptotic time $t=\infty$, but  does not satisfy the condition at $t \approx T_0$.   
 The state expressed by this wave function  does not ensure that represent an isolate state.  In  previous studies of the wave packet scattering \cite{Schiff,Iwanami-kouza,Messia}, the wave functions at $t= \infty$ were  studied, and cross section were computed from them. Because the total probability varies with time   at $t=T_0$, these wave functions do not express  isolated states.

\section{Summary}
 
Non-orthogonality of continuum stationary  states with different energies is an issue in  mathematics, 
and  has the important implication in physics of scattering. Requireing  physical conditions $H_1$:
physical quantities are  scaling invariant in large system, and $H_2$: physical system is isolated and is expressed by a state of  a constant norm, the  rigorous form of scalar products  are obtained using  overlap integrals of the  continuum states. Applying the integrals for the plane waves, Eqs. $(\ref{dirac_delta})$,  $(\ref{delta_m})$,  and 
$(\ref{dirac_delta_2})$ in Theorem 1, those of stationary states  Eqs.$(\ref{scalar_product})$ and $(\ref{non-conserving})$ in Theorem 2 are expressed by the reflection and transmission coefficients, $R(k)$ and $T(k)$.    

Questions 1 and  2 on the orthogonality of states with different energies in Introduction are answered as follows: 

$O_1$.  The stationary continuum  states with different energies  are not  orthogonal for most  short-range potentials. Superpositions of these states have  the net probability current. 

$O_2$. The stationary continuum  states with  different energies are orthogonal in a free system, a uniform force,  harmonic oscillator, Landau levels,  periodic potentials, and confining potentials in three dimensions. 
Superpositions of the  states with different energies do not have  net probability current. 
 
From overlap integrals, we found that wave packets  constructed from stationary continuum  states represent isolate states  in  potentials in $O_2$. The wave packets   have constant norms, and represent isolate states. These   provide   consistent  scattering amplitudes and probabilities in a time-independent formalism (1B).  On the other hand, the continuum states in $O_1$, i.e., short-range potentials, do not satisfy the orthogonality.  Wave packets constructed from superpositions of stationary states neither  have constant norms nor  represent isolate states. These  do not provide   consistent  scattering amplitudes and probabilities. In these systems, wave packets constructed from free waves satisfy  the orthogonality and their superpositions have constant norms.   Consequently, the wave packet amplitudes  derived from a time dependent formalism (2B) defined by perturbative expansion  in the interaction picture satisfy all the requirements. This result is in agreement with the completeness proof  of  eigensolutions by Newton, which was based on perturbative expansion of Green functions \cite{newton,newton_p,coulomb_c}. The nonorthogonality of rigorous continumm states with different energies  in short range potentials  provides a consistent view towards the sizes of wave packets in high energy physics, \cite{ishikawa-jinnouchi}, and are useful for computing exact  probabilities, \cite{ishikawa-ann}.

Conflict of interest statement: There is no disclosure to be made. 
Ethics statement: This article does not contain any studies involving human or animal participants.   
\subsection*{Acknowledgments}
 The author thanks Drs. K.-y. Oda, K. Nishiwaki,  R. Sato, and K. Seto  for useful discussions.  This work was partially supported by a 
Grant-in-Aid for Scientific Research ( Grant No. JP21H01107). 
 \appendix

\section{Non-orthogonality of continuum  states in the literature  }
The context of self-adjoint Hamiltonian has been fully clarified by Kato, Kuroda, Simons, and other mathematical physicists. The non-orthogonality of exact  solutions in solvable models would have been  known by them, but received little attention. This would be  because  the  cross section with stationary states, which were standard quantities in normal situations, can be computed  irrespective of non-orthogonality. We have reconsidered  these issues with respect to the recent interest in  the foundations of quantum mechanics. 

The Landau-Lifshitz textbook shows $ \langle \phi(k')|\phi(k) \rangle =\delta(E(k')-E(k)) + \delta $, where $\delta$ does not vanish at $E(k') \neq E(k)$. Nevertheless,   most studies have investigated physical quantities from probability flow  of $ |\phi(k) \rangle$. The flow  is irrelevant to the non-orthogonality of the states, and is not affected by    $\delta$. Details are provided  on  pages 62 and 63, especially on the fifth lines from the top of page 63 of \cite{Landau-Lifshits}.

Some works study   a finite volume system, in which periodic boundary condition in spaces are  considered. This works for  $\frac{d^2}{dx^2}$.  However,  variable $x$ is not periodic, and  incoming states differ  from outgoing states and do not satisfy the periodic boundary condition. \cite{Schiff}

The textbook ``Quantum mechanics, (Iwanami )'',  provides  a detailed account of  scatterings, and  uses regularized potential  $e^{-\epsilon |t|} V(x)$ in wave packet formalism.  This book helps compute  the wave function in the asymptotic region $\psi_{asy}(t,x)$ first, and such that a probability flux is obtained  and a cross section is computed.    Wave packet scattering  is studied within  an approximation in which  the square $|\psi(x)|^2$ is a probability at $x$  \cite{Iwanami-kouza}. In his comprehensive explanation  of the one-dimensional scattering, Lipkin states that the wave packet treatment is the most effective. However, there is  absolutely no detail account provided.  The orthogonality of the continuum  eigenfunctions of different  eigenvalues is known  to hold only in the leading order. See Page 231 for  more details. \cite{Lipkin}

Although textbooks on scatterings have provided almost complete descriptions,  the non-orthogonality of continuum stationary  states with different energies have not been addressed  \cite{goldberger,taylor,newton} \cite{Messia}.

\section{Regularization dependence}
The scalar products of  continuum  stationary states  defined  by a limit $\Lambda  \rightarrow \infty$ of those of  a region $ -\Lambda  \leq x \leq  \Lambda $  are compared with the scalar products  of the regularized functions.  Both methods give  similar results for the most singular term that is  proportional to the delta function.  The  next order terms are more involved because .the reduction of the integrals to the boundary terms is not exact in the latter method. We elucidate the  reason for these  differences.   The naive limit from $\Lambda  \rightarrow \infty$ gives reasonable behaviors and is  suitable for finding   non-orthogonality of the states.    
We focus on  cases that $|k_1-k_2|$ or $|k_1+k_2|$ are not large. 
\subsection{Free particle}

\subsubsection{Direct integral of the scalar product}

We define integrals, 
\begin{eqnarray}
I(x_1,x_2;k_1-k_2)&=&\int_{x_1}^{x_2} dx \phi(k_2,x)^{*} \phi(k_1,x) \\
 I^{2\epsilon}(x_1,x_2;k_1-k_2)&=&\int_{x_1}^{x_2} dx\phi(k_2,x)^{*} \phi(k_1,x)  e^{-2\epsilon |x|}.
\end{eqnarray}
For $\phi(k,x)=e^{ikx}, E(k)=\frac{1}{2} k^2$  These are computed as 
 \begin{eqnarray}
I(x_1,x_2;k_1-k_2)&=&\frac{e^{i(k_1-k_2)x_2}-e^{i(k_1-k_2)x_1}}{i(k_1-k_2)}\label{bulk_integral_1}\\
 I^{2\epsilon}(x_1,x_2;k_1-k_2)&=&\frac{e^{i(k_1-k_2+2i \epsilon)x_2}-e^{i(k_1-k_2+2i\epsilon)x_1}}{i(k_1-k_2+2i \epsilon)}. \label{bulk_integral_2}
\end{eqnarray} 

In a  limit , $x_2=\Lambda( \rightarrow \infty), x_1=L(finite) $
\begin{eqnarray}
I(L,\Lambda;k_1-k_2)
&= &e^{i(k_1-k_2)(L)/2} \frac{2i \sin (k_1-k_2) (\Lambda-L)/2}{i(k_1-k_2)} \nonumber \\
&\rightarrow &e^{i(k_1-k_2)(L)/2} \pi \delta( k_1-k_2).   
\end{eqnarray}

A square of an amplitude becomes 
\begin{eqnarray}
|e^{i(k_1-k_2)(\Lambda+L)/2} \frac{2i \sin (k_1-k_2) (\Lambda-L)/2}{i(k_1-k_2)}|^2=| \frac{2 \sin (k_1-k_2) (\Lambda-L)/2}{(k_1-k_2)}|^2,\nonumber \\
& &
\end{eqnarray}
and  the normalization integral is given by
\begin{eqnarray}
\int_{-\infty}^{\infty} d(k_1-k_2)| I(L,\Lambda,k_1-k_2)|^2=(\Lambda-L) \pi.
\end{eqnarray}

Next we study   $I^{2\epsilon}$. This is expressed in the limit $\Lambda \rightarrow \infty$ as
\begin{eqnarray}
I^{2\epsilon}(L,\Lambda;k_1-k_2)&=&\frac{e^{i(k_1-k_2+2i \epsilon)\Lambda}-e^{i(k_1-k_2+2i \epsilon)L}}{i(k_1-k_2+2i \epsilon)} \nonumber \\
&\rightarrow&\frac{ie^{i(k_1-k_2+i 2\epsilon)L}}{k_1-k_2+i 2\epsilon}\nonumber \\
& &=e^{i(k_1-k_2+i2\epsilon)L}(\pi\delta(k_1-k_2)+i\frac{k_1-k_2}{(k_1-k_2)^2+4\epsilon^2} ). \nonumber \\
& &
\end{eqnarray} 
A square of the modulus of the right-hand side is independent of an overall phase factor and 
\begin{eqnarray}
|\frac{ie^{i(k_1-k_2+2\epsilon)L}}{k_1-k_2+i 2\epsilon}|^2=e^{-4\epsilon L}\frac{1}{(k_1-k_2)^2+4\epsilon^2} \nonumber\\
& &
\end{eqnarray}
 has a contribution from the second term. The function in a small $|k_1-k_2|$ region  behaves like the Dirac delta function of $k_1-k_2$.
\begin{eqnarray}
\int_{-\infty}^{\infty}d(k_1-k_2) |I_{2\epsilon}(0,\infty,k_1-k_2)|^2=\frac{1}{2 \epsilon} \pi.
\end{eqnarray}
  
Comparison of asymptotic limit (2); $x_2=\Lambda( \rightarrow \infty), x_1=0$
\begin{eqnarray}
I(0,\Lambda;k_1-k_2)&=&\frac{e^{i(k_1-k_2)\Lambda }-1}{i(k_1-k_2)}\nonumber \\
&= &e^{i(k_1-k_2)\Lambda/2} \frac{2i \sin (k_1-k_2) \Lambda/2}{i(k_1-k_2)} \nonumber \\
&\rightarrow & \pi \delta( k_1-k_2),  \\ 
I^{2\epsilon} (0,\Lambda;k_1-k_2)&=&\frac{e^{i(k_1-k_2+2i \epsilon)\Lambda}-1}{i(k_1-k_2+2i \epsilon)} \nonumber \\
&\rightarrow&\frac{i}{k_1-k_2+i 2\epsilon}=\pi\delta(k_1-k_2)+i\frac{k_1-k_2}{(k_1-k_2)^2+4\epsilon^2}. \nonumber \\
& &
\end{eqnarray} 

Comparisons of $I(0,\Lambda;k_1-k_2) $ and $I^{2\epsilon}(0,\infty ;k_1-k_2) $:

(1) ~The leading terms expressed by the delta function agree and are expressed by $\delta(k_1-k_2)$.

(2)~ Integrals of square $| I(L,\Lambda,k_1-k_2)|^2$ is $\pi (\Lambda-L)$, which is proportional to the spatial length.  The one of  $|I_{2\epsilon}(0,\infty;k_1-k_2) |^2$ is $\pi \frac{1}{2\epsilon}$, which is  proportional to the regularization parameter.  $|I(L,\Lambda;k_1-k_2))|^2$  and $|I^{2\epsilon}(0,\infty;k_1-k_2)|^2$ are sloghtly different.  Mathematically $I^{2\epsilon}(0,\infty;k_1-k_2)$ is simpler than $I(L,\Lambda;k_1-k_2)$ but for a physical application, $\frac{1}{\epsilon}$ is replaced with a physical scale at the end of calculations. Because $I(L,\Lambda;k_1-k_2)$ has a physical  scale from its definition, this gives clearer meaning  than the former one when compared with experiments. 

(3)~ The next order terms are different.   $I(0,\infty;k_1-k_2)$ is simply proportional to the delta function, and reveals a manifest orthogonality of a free particle. 
\subsubsection{Reduction to boundary terms in free waves}
The scalar product of stationary states is reduced to the boundary values, but  
those of regularized functions are slightly modified  to disagree with the value of exact eigenstates.
 Validity of the  reduction of the integral is checked.

 Using regularized  free waves,
$ \psi(k,x)=e^{ikx-\epsilon x}, \psi(k,x)^{*}=e^{-ikx-\epsilon x}$ the boundary contributions are written as follows:
\begin{eqnarray}
& &[d \psi(k_2,x)^{*} \psi(k_1)-\psi(k_2,x)^{*} d \psi(k_1)]|_{x_1}^{x_2} \nonumber \\
& &=-i(k_2-i\epsilon )-i(k_1+i\epsilon)(e^{i(k_2-k_1+2\epsilon)x_2}-e^{i(k_2-k_1+2\epsilon)x_1}) \nonumber \\
& &=(-i)(k_1+k_2)(e^{-i(k_2-k_1+2\epsilon)x_2}-e^{-i(k_2-k_1+2\epsilon)x_1}) 
\end{eqnarray}
and $E(k)=(k+i\epsilon)^2/2$,
an equality for two expressions is 
\begin{eqnarray}
& &I^{2\epsilon}=\frac{1}{2(E(k_2)^{*}-E(k_1)) }(-i)(k_1+k_2)(e^{-i(k_1-k_2+2i \epsilon)x_2}-e^{-i(k_1-k_2+2i\epsilon)x_1}) \nonumber \\
& &=\frac{e^{i(k_1-k_2+2i \epsilon)x_2}-e^{i(k_1-k_2+2i\epsilon)x_1}}{i(k_1-k_2+2i \epsilon)}.\label{boundary_integral_2}
\end{eqnarray}
Eq.$(\ref{boundary_integral_2})$ agrees with Eq.$(\ref{bulk_integral_2})$.


\subsection{Reduction of integrals in  solvable potentials }
 Solvable  potentials are studied.  

1.The wave function in a square well potential in $x< -a$ is
\begin{eqnarray}
\psi(k_l)=e^{ik_l x}+R(k_l) e^{-ik_l x}, \psi(k_l)^{*}=e^{-ik_l x}+R(k_l) e^{ik_l x}. 
\end{eqnarray}

(1) Direct integral without damping factor. 
\begin{eqnarray}
& &I(x_1,x_2;k_1-k_2)=\frac{e^{-i(k_2-k_1)x_2}-e^{-i(k_2-k_1)x_1}}{-i(k_2-k_1)}+R(k_1)\frac{e^{-i(k_2+k_1)x_2}-e^{-i(k_1+k_2)x_1}}{-i(k_1+k_2)} \nonumber\\
& &+R(k_2)^{*}\frac{e^{i(k_1+k_2)x_2}-e^{i(k_1+k_2)x_1}}{i(k_1+k_2)}+R(k_2)^{*}R(k_1) \frac{e^{i(k_2-k_1)x_2}-e^{i(k_2-k_1)x_1}}{i(k_2-k_1)}.\nonumber \\
& &
\end{eqnarray}

 (2) Reduction to the boundary without damping factor is
\begin{eqnarray}
2(E(k_2)-E(k_1)) \int_{x_1}^{x_2} \psi(k_2,x)^{*} \psi(k_1,x)=-(d \psi(k_2,x))^{*} \psi( k_1,x)-( \psi(k_2,x))^{*} d \psi( k_1,x)|_{x_1}^{x_2}, \nonumber \\
& & \label{bulk_boundary}
\end{eqnarray}
where 
\begin{eqnarray}
& &(d \psi(k_2,x))^{*} \psi( k_1,x)-( \psi(k_2,x))^{*} d \psi( k_1,x)  \nonumber\\
& &=-i(k_1+k_2)( e^{-i(k_2-k_1)x_2} -e^{-i(k_2-k_1)x_1})  -i(k_2-k_1)R(k_2)( e^{-i(k_1+k_2)x_2} -e^{-i(k_1+k_2)x_1}) \nonumber \\
& &+i(k_2-k_1)R(k_2)^{*}(e^{i(k_1+k_2)x_2} -e^{i(k_1+k_2)x_1})+i(k_1+k_2)R(k_2)^{*}R(k_1)( e^{i(k_2-k_1)x_2}- e^{i(k_2-k_1)x_1} ).\nonumber \\
\end{eqnarray}
Both  agree.  The integral is converted to the surface term.  

(3) Boundary reduction with  regularized functions.

Regularized functions : $(x<0)$
\begin{eqnarray}
& & \psi_r(k,x)=e^{ikx+\epsilon x}+R(k_l)e^{-ik_lx+\epsilon x}, \psi_r(k,x)^{*}=e^{-ikx+\epsilon x}+R(k_l)^{*}e^{ik_lx+\epsilon x} \nonumber \\
& &d^2 \psi_r(k,x)=(ik +\epsilon)^2e^{ikx+\epsilon x}+R(k_l)(-ik+\epsilon)^2 e^{-ik_lx+\epsilon x}  \nonumber \\
& &= (-ik +\epsilon)^2\psi_r(k,x)  +4i k \epsilon e^{ikx+\epsilon x}  \neq E(k) \psi_r(k,x)   \label{derivative_1}\\
& &d^2\psi_r(k,x)^{*}=(-ik+\epsilon)^2e^{-ikx+\epsilon x}+R(k_l)^{*}(ik+\epsilon)^2e^{ik_lx+\epsilon x} \nonumber \\
& &=(ik+\epsilon)^2 \psi_r(k,x)^{*} -4i \epsilon e^{-ikx+\epsilon}
\neq E(k)^{*} \psi(k,x)^{*}.  \label{derivative_2}
\end{eqnarray}
Due to the damping factor $ i\epsilon$ in Eqs.$(\ref{derivative_1})$ and $(\ref{derivative_2})$, a square of the momentum $k-i\epsilon$ of the right-moving wave are not the same  as the square of  left-moving wave 
 $-k-i\epsilon$. These waves are not exact  eigenstates of the Hamiltonian, and the integral is not  reduced to the boundary terms. 
The deviation  $2i\epsilon k$ is small but depends on the momentum $p$. Its impact  is sensitive issue,  but Eq.$(\ref{bulk_boundary})$ is not satisfied.

This is  also revealed by a direct calculation,   
\begin{eqnarray}
& &I^{2\epsilon}(k_1-k_2)=\int_{x_1}^{x_2} dx \psi_r(k_2,x)^{*} \psi_r(k_1,x) \\
& &=\frac{e^{-i(k_2-k_1)x_2 +2 \epsilon x_2}-e^{-i(k_2-k_1)x_1+2 \epsilon x_1}}{-i(k_2-k_1)-i2 \epsilon}+R(k_1)\frac{e^{-i(k_1+k_2)x_2+2 \epsilon x}-e^{-i(k_1+k_2)x_1 +2 \epsilon x}}{-i(k_1+k_2)+2 \epsilon} \nonumber\\
& &+R(k_2)^{*}\frac{e^{i(k_1+k_2)x_2 +2 \epsilon x_2}-e^{i(k_1+k_2)x_1+2 \epsilon x_1 }}{i(k_1+k_2)+2 \epsilon }+R(k_2)^{*}R(k_1) \frac{e^{i(k_2-k_1)x_2+2 \epsilon x_2 }-e^{i(k_2-k_1)x_1 +2 \epsilon x_1}}{i(k_2-k_1) +2 \epsilon}. \nonumber \\
& &
\end{eqnarray}
This cannot be written by a reduced boundary term. 
\begin{eqnarray}
& &2(E(k_2)^{*}-E(k_1)) \int_{x_1}^{x_2} \psi_r(k_2,x)^{*} \psi_r(k_2,x)\nonumber \\
& &\neq -(d \psi_r(k_2,x))^{*} \psi_r( k_1,x)-( \psi_r(k_2,x))^{*} d \psi_r( k_1,x)|_{x_1}^{x_2}, 
\end{eqnarray}
where 
\begin{eqnarray}
& &E(k_2)^{*}-E(k_1)=\frac{1}{2}(k_1+k_2)(k_2-k_1 +2 i \epsilon).
\end{eqnarray}

2.$V_0 \frac{1}{\cosh^2 \mu x}$

The integrals of regularized functions are not reduced to the boundary terms.  

\subsubsection{ Waves in finite systems of boundary conditions }
One particle state  in a  system of  finite size is expressed by a complete set of functions which are characterized by a boundary condition. In  periodic boudnary condition, an out-going state  is equivalent to an in-coming state. These states do not express 
 scattering  phenomena in which     an out-going state is different from an in-coming state.  Accordingly, general boundary condition is studied.

{\bf Periodic boundary condition}

Under the periodic boundary condition $\phi(x+2\Lambda)=\phi(x)$,
\begin{eqnarray}
& & e^{ik_i(x+2 \Lambda)}= e^{ik_i(x)}, 2 k_i \Lambda =2\pi n_i \nonumber \\
& & k_i~=\frac{ \pi}{\Lambda} n_i.
\end{eqnarray}
Then 
\begin{eqnarray}
& &  \int_{-\Lambda}^{\Lambda } dx e^{-i(k_2-k_1)x} = \frac{e^{-i(k_2-k_1)\Lambda} -e^{i(k_2-k_1)\Lambda }}{-i(k_2-k_1)}=0, \nonumber \\
& &
\end{eqnarray}
for  $k_2-k_1 \neq 0$.

{\bf Twisted  boundary condition}

Under the  twisted boundary condition  $\phi(x+2\Lambda)=\phi(x)e^{i\theta}$,
\begin{eqnarray}
& & e^{ik_1(x+2 \Lambda)}= e^{ik_1(x)} e^{i\theta_1}, 2 k_1 \Lambda =2\pi n_1 +\theta_1  \nonumber \\
& & k_1~=\frac{ \pi}{\Lambda} n_1+\frac{\theta_1}{2 \Lambda},
\end{eqnarray}
and
\begin{eqnarray}
& & e^{ik_2(x+2 \Lambda)}= e^{ik_2(x)} e^{i\theta_2}, 2 k_2 \lambda =2\pi n_2 +\theta_2  \nonumber \\
& & k_2~=\frac{ \pi}{\Lambda} n_2+\frac{\theta_2}{2 \Lambda},
\end{eqnarray}
then 
\begin{eqnarray}
& &  \int_{-\Lambda}^{\Lambda } dx e^{-i(k_2-k_1)x} = \frac{e^{-i(\theta_2-\theta_1)/2} -e^{i(\theta_2-\theta_1)/2 }}{-i(k_2-k_1)}, \nonumber \\
& &= \Lambda \frac{-2i \sin(\frac{(\theta_2-\theta_1)}{2})} {-i({\pi}(n_2-n_1) +\frac{\theta_2-\theta_1}{2})}, 
\end{eqnarray}


\end{document}